\documentclass{aa}
\usepackage{graphicx}
\usepackage{mathtools}
\usepackage{amsmath}
\usepackage{fancyref}
\usepackage{longtable}
\usepackage{rotating}
\usepackage{pdflscape}
\usepackage{enumerate}
\usepackage[breaklinks=false, colorlinks=true, citecolor=blue,urlcolor=blue,filecolor=blue, linkcolor=blue]{hyperref}
\usepackage{txfonts}
\usepackage{natbib}
\begin{document}

   \title{Search for associations containing young stars (SACY)}

   \subtitle{VII. New stellar and substellar candidate members in the young associations\thanks{Based on FEROS observations obtained during CNTAC programme CN2015B-9 and observations made with the HERMES spectrograph mounted on 
the 1.2\,m Mercator Telescope at the Spanish Observatorio del Roque de los Muchachos 
of the Instituto de Astrof\'isica de Canarias.}$^{,}$\thanks{Appendices~\ref{sec:notes_individ_sources}, \ref{sec:pre_identified_wide_comps}, \ref{sec:app_hr_diagrams}, and \ref{sec:orig_sample} are available in electronic form at \url{http://www.aanda.org}.  Table~\ref{tab:online_appendix_tab} is only available at the CDS via anonymous ftp to \nolinkurl{http://cdsarc.u-strasbg.fr (ftp://123.45.678.9)} or via \url{http://cdsarc.u-strasbg.fr/viz-bin/qcat?J/A+A/XXX/XXX}}}

  \author{P. Elliott
          \inst{1, 2}\and
          A. Bayo
          \inst{3}\and
          C. H. F. Melo
          \inst{1}\and
          C. A. O. Torres\inst{4}
          \and 
          M. F. Sterzik\inst{5}
          \and
          G. R. Quast\inst{4}     
          \and
          D. Montes\inst{6} 
          \and
          R. Brahm\inst{7, 8} 
          }
   \institute{European Southern Observatory, Alonso de Cordova 3107, Vitacura Casilla 19001, Santiago 19, Chile
              \\
              \email{pe210@exeter.ac.uk}
         \and
            School of Physics, University of Exeter, Stocker Road, Exeter, EX4 4QL
        \and 
           Departamento de F\'isica y Astronom\'ia, Facultad de Ciencias, Universidad de Valpara\'iso, Av. Gran       Breta\~na 1111, 5030 Casilla, Valpara\'iso, Chile
         \and
            Laborat\'orio Nacional de Astrof\'isica/ MCT, Rua Estados Unidos 154, 37504-364 Itajub\'a (MG), Brazil
            \and
            European Southern Observatory, Karl-Schwarzschild-Str. 2, D-85748 Garching, Germany
            \and
            Departamento de Astrof\'isica y Ciencias de la Atm\'osfera, Facultad de Ciencias F\'isicas, Universidad Complutense de Madrid, 28040 Madrid, Spain
            \and
            Instituto de Astrof\'isica, Facultad de F\'isica, Pontificia Universidad Cat\'olica de Chile, Av. Vicu\~na Mackenna 4860, 7820436 Macul, Santiago, Chile
        \and
Millennium Institute of Astrophysics, Av. Vicu\~na Mackenna 4860, 7820436 Macul, Santiago, Chile
             }

   \date{Received 4\,February 2016; accepted 7\,March 2016}

  \abstract
   {The young associations offer us one of the best opportunities to study the properties of young stellar and substellar objects and to directly image planets thanks to their proximity ($<$200\,pc) and age ($\approx$5-150\,Myr). However, many previous works have been limited to identifying the brighter, more active members ($\approx$1\,M$_\odot$) owing to photometric survey sensitivities limiting the detections of lower mass objects.}
   {We search the field of view of 542 previously identified members of the young associations to identify wide or extremely wide (1000-100,000\,au in physical separation) companions.}
   {We  combined 2MASS near-infrared photometry ($J$, $H$, $K$) with proper motion values (from UCAC4, PPMXL, NOMAD) to identify companions in the field of view of known members.  We collated further photometry and spectroscopy from the literature and conducted our own high-resolution spectroscopic observations for a subsample of candidate members. This complementary information allowed us to assess the efficiency of our method.}
   {We identified 84 targets (45:\,0.2-1.3\,M$_\odot$, 17:\,0.08-0.2\,M$_\odot$, 22:\,$<$0.08\,M$_\odot$) in our analysis, ten of which have been identified from spectroscopic analysis in previous young association works.  For 33 of these 84, we were able to further assess their membership using a variety of properties (X-ray emission, UV excess, H$_\alpha$, lithium  and K\,I equivalent widths, radial velocities, and CaH indices).  We derive a success rate of 76--88\%  for this technique based on the consistency of these properties.}
   {Once confirmed, the targets identified in this work would significantly improve our knowledge of the lower mass end of the young associations.  Additionally, these targets would make an ideal new sample for the identification and study of planets around nearby young stars.  Given the predicted substellar mass of the majority of these new candidate members and their proximity, high-contrast imaging techniques would facilitate the search for new low-mass planets. 
}


   \maketitle
%

\section{Introduction}
\label{sec:intro}

The very first members of the TW Hydrae association, one of the youngest ($\approx$10\,Myr) and closest ($\approx$50\,pc) young associations, were identified more than two and a half decades ago \citep{delaReza1989}. 
The authors noted that a handful of isolated T Tauri stars with high Galactic latitudes shared similar Galactic kinematics.

These young stars have been identified using a variety of properties (X-ray emission, lithium absorption, H$_\alpha$ emission, common Galactic kinematics, visible, and near-infrared photometry) to identify co-moving groups and distinguish them from older field stars.  However, a common property of much of these works was the identification of stars with spectral types earlier than $\approx$K0.  This was the result of the sensitivities of surveys, such as HIPPARCOS \citep{Perryman1997} and the ROSAT all-sky survey \citep[RASS;][]{Voges1999}, which were used to build the samples.  Recently there have been many works focussed on identifying the lower mass members of the young associations e.g. \cite{Shkolnik2012, Schlieder2012, Rodriguez2013, Malo2014, Kraus2014, Gagne2015, Binks2015}.  Again, these surveys use a variety of techniques, however, the initial samples of low-mass candidate members are usually built upon multiple-catalogue photometry combined with proper motions. The candidates are then followed up using mid or high resolution to produce radial velocities (RVs) and signatures of youth, such as H$_\alpha$ emission and lithium absorption.  As a result of the optical magnitudes of these targets (typically $V$ $>14$\,mag.) the observations are very time expensive, therefore refinement of the initial sample is crucial where possible.

The identification of further low-mass members in the young associations is extremely important for a number of reasons.  These targets offer us one of the best opportunities to image exoplanets using high-contrast imaging.  The dimmer host star in addition to a hotter planet (because of to its youth) significantly reduces contrast constraints \citep{Chauvin2015, Bowler2013}.  Additionally, the young associations currently appear to be very incomplete for low-mass stars.  This incompleteness means the initial mass function is very poorly constrained below $\approx$1\,M$_\odot$.  Therefore, no meaningful comparison can be made between the IMFs of young associations, formed in low-density regions, and those of higher density clusters.

The work presented here forms part of larger project to study multiplicity in the young associations \citep{Elliott2014, Elliott2015}. These works have used the search for associations containing young stars (SACY) dataset.  Stars are defined as members of distinct associations in this dataset by the convergence method \citep{Torres2006}. In this work we extend the membership list including high-probability targets identified in other works; see Section~\ref{sec:sample} for details. The targets identified in this work are candidate wide companions to these previously identified high-probability members.  In that respect this work differs from the majority of previous works aimed at identifying new members and is more similar to works related to wide binaries \citep{Caballero2009, Caballero2010, Alonso-Floriano2015}.  The formation mechanism and implications of such wide (1000-100,000\,au) companions will be discussed in further work.  Additionally detection limits derived from the analysis presented here will form part of another study combining radial velocities, adaptive optics (AO)-imaging and 2MASS photometry to derive multiplicity statistics across 8 orders of magnitude in physical separation in the young associations.  In this work we present the technique we employed to identify the wide companion candidates, an evaluation of our method (using available literature and our own high-resolution spectroscopy observations) and the list of candidate members with their current membership status.

The manuscript is arranged as follows.  Section~\ref{sec:sample} explains how the sample for this work was formed.  Section~\ref{sec:technique} details our technique for identifying new members.  Section~\ref{sec:further_info} outlines the sources where we gathered additional information. Section~\ref{sec:new_spec} describes the high-resolution spectra we analysed for a subsample of identified targets.  Section~\ref{sec:cand_properties} discusses the properties of the candidates we identified. Section~\ref{sec:conclusions} concludes this work evaluating the success of the technique. 

\section{The sample}
\label{sec:sample}

We collated high-probability members from a collection of prominent moving group works to date \citep{Torres2008, Zuckerman2011, Malo2014,Kraus2014, Elliott2014, Murphy2015}, which cover a wide range of spectral types (B2-M5) and ages ($\sim$5-150\,Myr), totalling 542 targets in nine associations. A summary of the associations can be found in Table~\ref{tab:ass_summary}.  We did not put any constraints on the mass of the known members as we want to maximise our sample size.  However, we are limited in sensitivity to wide companions depending on the position of the source on the sky, its proper motion, and density of objects in the field of view (FoV), as discussed in Section~\ref{sec:technique}. The targets that make up the sample and their basic properties are listed in Table~\ref{tab:orig_sample}.

{
\begin{table}
\tiny
\caption{Summary of the associations studied in this work.}
\begin{tabular}{p{2cm} p{1.1cm} p{1cm} p{1cm}  p{1cm}}
\hline\hline\\
  \multicolumn{1}{l}{Ass.} &
  \multicolumn{1}{l}{Ass. ID} &
  \multicolumn{1}{l}{Age} &
  \multicolumn{1}{l}{Age ref.} &  
  \multicolumn{1}{l}{No. of objects} \\[1ex]
\hline\\
AB Dor & ABD & 100-150 & 2,7 & 105\\
Argus  & ARG & 40-44 & 1,7 & 77\\
$\beta$-Pic  & BPC & 21-26 & 1,3,4,7 & 69 \\
Carina  & CAR & 35-45 & 1,3,7 & 48\\
Columba  & COL & 35-42 & 1,3,7 & 78\\
$\epsilon$-Cha  & ECH & 5-10 & 1,5,7 & 24\\
Octans  & OCT & 30-40 & 1,6,7 & 62\\
Tuc-Hor  & THA & 33-45 & 1,3,7 & 56\\
TW Hydrae  & TWA & 10-12 & 1,3,7 & 23\\
\hline\\[-0.1ex]
\end{tabular}
{\footnotesize \bf References.}{\footnotesize~1: \cite{Torres2006}, 2: \cite{daSilva2009}, 3: \cite{Bell2015}, 4: \cite{Binks2014}, 5: \cite{Murphy2013}, 6: \cite{Murphy2015}, 7: Torres et al. in prep.}
\label{tab:ass_summary}
\end{table}}

\section{Searching for new members as very wide companions}
\label{sec:technique}

To identify very wide companions we used the point source catalogue of 2MASS \citep{Cutri2003} for near-infrared photometry ($J$, $H$, $K$).  We combined this with proper motions from the naval observatory merged astrometric dataset (NOMAD), the proper motion extended catalogue (PPMXL), and   the United States naval observatory (USNO) CCD astrograph catalogue (UCAC4) \citep[][respectively]{Zacharias2005,Roeser2010,Zacharias2012}.

One aim of this search was to produce robust detection limits of our entire sample out to 100,000\,au in physical separation from the known member.  These limits will be used in further work as explained in Section~\ref{sec:intro}.  We first calculated the angular query needed to achieve a physical search radius out to 100,000\,au using the distance to the source (either from parallax \citealt{ESA1997, vanLeeuwen2007} or derived kinematically using the convergence method \citealt{Torres2006}). We then converted the apparent magnitude of any sources in the FoV into absolute magnitude, assuming the distance of the known member. Using the available photometry ($J$, $H$, $K$) we constructed colour-magnitude diagrams ($H$-$K$, $K$ and $J$-$K$, $K$) for each FoV. We used the evolutionary tracks of \cite{Baraffe2015}, with the age of the association that the member belongs to.  We then used the photometry with derived statistical uncertainties and kinematic criteria (detailed below) to classify potential companions.
\subsection{Computing statistical photometric uncertainties}

The three photometric bands of 2MASS ($J$, $H$, $K$) do not offer a very wide range of colours to distinguish young objects from contamination.  Additionally proper motion values are only projected motions and therefore the distance to the object can be a severe limiting factor. For those reasons we took the time to produce photometric criteria that would limit the number of false-positives whilst attempting to minimise the possibility of true companion rejection.

First of all, for the
entire sample we computed the standard deviation of the difference between the colour of the isochrone and 2MASS photometry for each object.  The values were $\sigma_{\mathrm{H-K}}$=0.06\,mag and $\sigma_{\mathrm{J-K}}$=0.16\,mag for the colours $H$-$K$ and $J$-K, respectively.  This provided an estimate of the difference between the models and observations.  We used the generous criteria of 2$\sigma$ for both colours, however, the source must agree in both of these colours to be classified as a potential companion.

The colour and magnitude are related quantities and therefore an offset in magnitude can dramatically affect the agreement in colour.  An unresolved (equal-mass) companion can affect the magnitude of a source by up to $\approx$0.8\,mag.  We therefore adopted this value as our upper uncertainty on the absolute magnitude.  The effect of unresolved companions is asymmetric with respect to the magnitude, i.e. it only ever increases this value.  For the lower limit, we computed the average difference (0.15\,mag.) between the isochrone and the known member; in the cases that the known member was fainter than its isochrone,  we used 2$\sigma$ as our criterion again.

 We performed the analysis on a subset of the primary sources in our sample to test the effectiveness of our derived criteria.  Of 198 primaries, 188 were initially identified correctly.  Of the ten sources that were not identified, seven were known sources with two or more components in their photometry \citep[such as HD 217379, identified as a triple-lined spectroscopic binary in][]{Elliott2014}.  These are rare cases, however we do not want to exclude any sources (companions also have a small probability of being in an SB3 configuration) and therefore we increased the upper limit on the magnitude (1.2\,mag.).  With this new criterion we identified 195 / 198 of our primaries. Of the three sources (T Cha, CP-681894, TWA 22) that were still not identified, two have near-infrared photometric excesses (T Cha is a well-known disc system \citep{Weintraub1990} and CP-681894 shows strong H$_\alpha$ emission and K-band excess, indicative of accretion \citep{Schultz2005a}.  The other target, TWA 22 is a very tight, near equal-mass binary system \citep{Bonnefoy2009}.  Using the resolved photometry this system would have been classified correctly.  However, with the unresolved 2MASS photometry the system appears too red for its magnitude.  Owing to its proximity ($\approx$18\,pc) and its age (10-20\,Myr) the effect of the unresolved companion is exaggerated, however such young, nearby, low-mass (system mass: 220$\pm$21\,M$_\mathrm{Jup.}$) systems are very rare in our sample. Therefore, we proceeded with the criteria described above.

\begin{figure}[h]
\begin{center}
\includegraphics[width=0.45\textwidth]{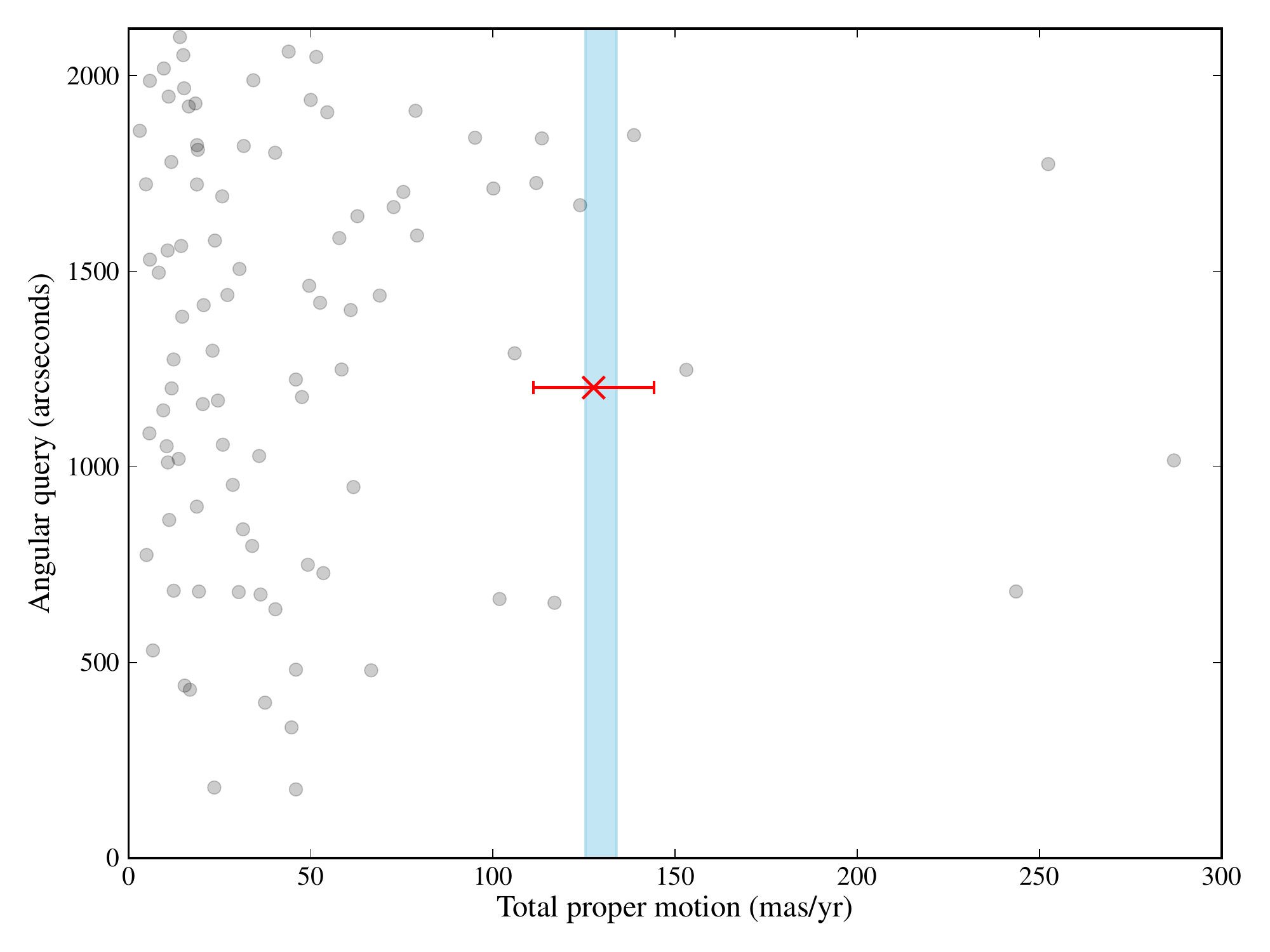}
\vspace{-0.2cm}
\caption{Graphical demonstration of our contamination / detection method for target Kap Psc (distance: 50\,pc).  The red markers indicate proper motion compatible sources (with 3$\sigma$ uncertainties) and the grey markers indicate incompatible sources. The blue shaded space is the 3$\sigma$ boundary of the proper motion of the primary target. The extent of the angular query is 2000\arcsec, equivalent to 100,000\,au at 50\,pc.}
\label{fig:contamination}
\end{center}
\end{figure}


This may induce a small bias in the way that we search for our companions, i.e. we could potentially exclude sources with strong infrared excess. However, only 3 / 198 were not classified because of their excess (1.5\%). Additionally the frequency of primordial discs for the ages of the young associations is very low.  For example, at 5\,Myr the frequency is $\approx$10\,\% and for populations older than 10\,Myr $<$5\,\%, \citep[see Figure~1 of][]{Mamajek2009}.   Furthermore this effect is less and less prominent as the mass of the companion decreases.  \cite{Andrews2013} showed that, for a sample of stars with spectral types earlier than M8.5 in Taurus, the disc mass decreases with the stellar mass.  If we consider this in terms of infrared photometry the excess scales with the luminosity of the object, i.e. a significant excess, considering the same detection threshold, is less likely for a lower mass star. Furthermore, if we assume a flat-mass ratio, as observed in many regions \citep[including close, $\approx$10-1000\,au, visual binaries in young associations;][]{Elliott2015} this bias becomes even less significant. 

\subsection{Using near-infrared photometry and proper motions to find wide companions}

Any source that met the criteria described above was classified as a potential companion within the FoV.  We then cross-matched all of these sources with the catalogues of UCAC4, PPMXL, and NOMAD, in that order of preference based on uncertainties, completeness, and catalogue-catalogue agreement, to collect the proper motion values.  

Finding physical extremely wide companions is limited by the projected motion of the target on the sky.  In most cases we do not have access to parallax or radial velocity values; we only have access to proper motion, which is intrinsically linked to its distance. Additionally the Galactic position of the target is a limiting factor, i.e. targets close to the Galactic plane (and close to the Sun) are likely to have a high number of sources in their FoV; this increases the false-positive probability even with the imposed photometric constraints.  Our targets are of different ages, at different distances and spatial positions and therefore there are many competing factors in estimating the contribution of contamination.  We treated the contamination of each source individually to find potential companions and also to derive the maximum angular separation space that we could probe. This method provides us with detection limits in angular separation, which is crucial for our broader statistical analysis to be presented in future work (Elliott et al. in prep.).

For sources with low or intermediate proper motion ($<$20\,mas) the chance of contamination increases significantly. In many works authors decide to use a hard cut-off (i.e. only considering sources with proper motions above a certain, somewhat arbitrary, threshold). However, information on the population can be lost as a source with intermediate proper motion is not analysed at all.  Below is a description of the method we employed to maximise the physical separation space we can query without significant contamination.

Figure~\ref{fig:contamination} graphically demonstrates our method for a primary target in our sample. The total proper motion, the x-axis, is defined as $\sqrt{\mu_{\alpha, *}^2+\mu_{\delta}^2}$.  Therefore, two sources may share the same total proper motion, however, not the same angle of projected motion.  It is this angle that distinguishes proper motion-compatible sources (red markers) from incompatible sources (grey markers).

All markers in Figure~\ref{fig:contamination} are sources with compatible 2MASS photometry. The grey circles show sources with an incompatible proper motion vector ($\geq 3 \sigma$ discrepancy), the red marker (at $\approx$1200\arcsec, 57,000\,au) represents sources with a compatible proper motion vector, and the blue shaded area is 3$\sigma$ proper motion space of the primary target. The total queried FoV is equivalent to 100,000\,au.  We use the point at which the total proper motion of the primary crosses any region in the parameter space with 2 or more incompatible sources as our maximum angular separation.  This is a very conservative estimate but proved very powerful. If the total proper motion of the primary never crosses any such regions we set our sensitivity to a radial physical distance of 100,000\,au, as in Figure~\ref{fig:contamination}.

\begin{figure}
\begin{center}
\includegraphics[width=0.45\textwidth]{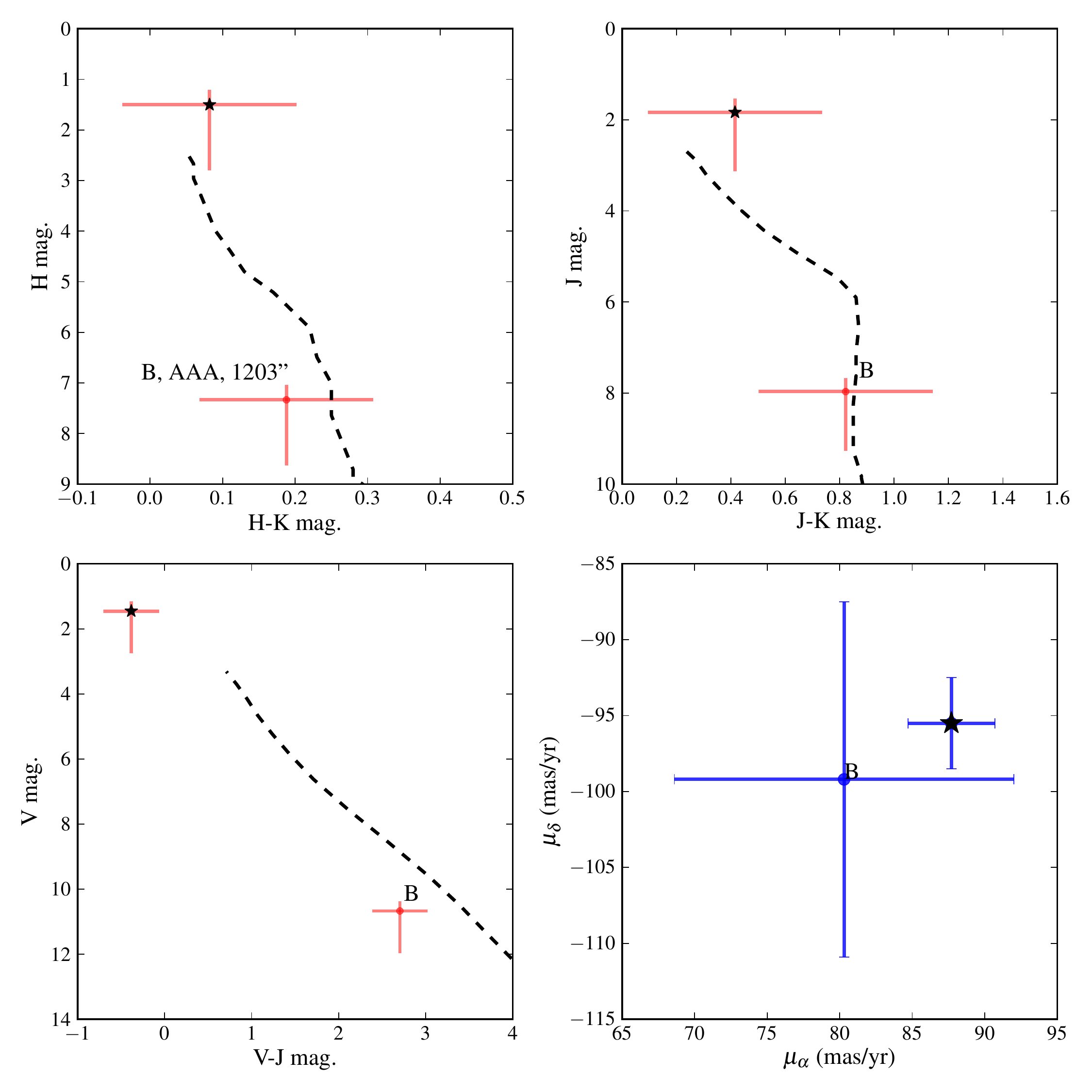}
\vspace{-0.2cm}
\caption{{\it Top left, top right, and bottom left panels:} colour-magnitude diagrams for Kap Psc+2MASS J23270114+0055200 in ($H$-$K$, $K$), ($J$-$K$, $K$), and ($V$-$J$, $V$), respectively.  The text next to the marker indicates the component designation, the quality flags of the 2MASS photometry and angular separation from the primary.  The dotted lines are the evolutionary tracks of \cite{Baraffe2015} using the age of the moving group. {\it Bottom right panel}: Proper motion values of the primary and companion.}
\label{fig:hr_example}
\end{center}
\end{figure}

Figure~\ref{fig:hr_example} shows a summary of the photometry ($V$, $J$, $H$, $K$) and proper motion of the potential companion from Figure~\ref{fig:contamination}.  The agreement in the colour $V$-$K$ is generally very poor and is discussed more thoroughly in Section~\ref{subsec:vmag_disc}.
For the inner angular query we use the limit given by the PSF of 2MASS ($\approx$3\arcsec) that relates to a physical inner limit of 18--570\,au for our nearest and furthest sources, respectively. 

\section{Further information from catalogues and literature}
\label{sec:further_info}

Table~\ref{tab:prev_surveys} summarises the catalogues and works with which we cross-matched our sample to gain further photometry, spectroscopy, kinematics, and information of multiplicity.  Where the cross-match was successful we collated any available notes of the source from the works.

We also compared our results to other works that have focussed on very wide companions to stars in the moving groups \citep{Kastner2011, Kastner2012, Scholz2005, Looper2010, Feigelson2006, Alonso-Floriano2015}.  A detailed comparison of these targets can be found in Appendix~\ref{sec:pre_identified_wide_comps}. Additionally we queried Simbad \citep{Wenger2000} for each object to collate any noteworthy features, analysis, and further parameters available.  

\subsection{Photometry from the Catalina survey}
\label{sec:catalina_phot}

\begin{figure}[h]
\begin{center}
\includegraphics[width=0.49\textwidth]{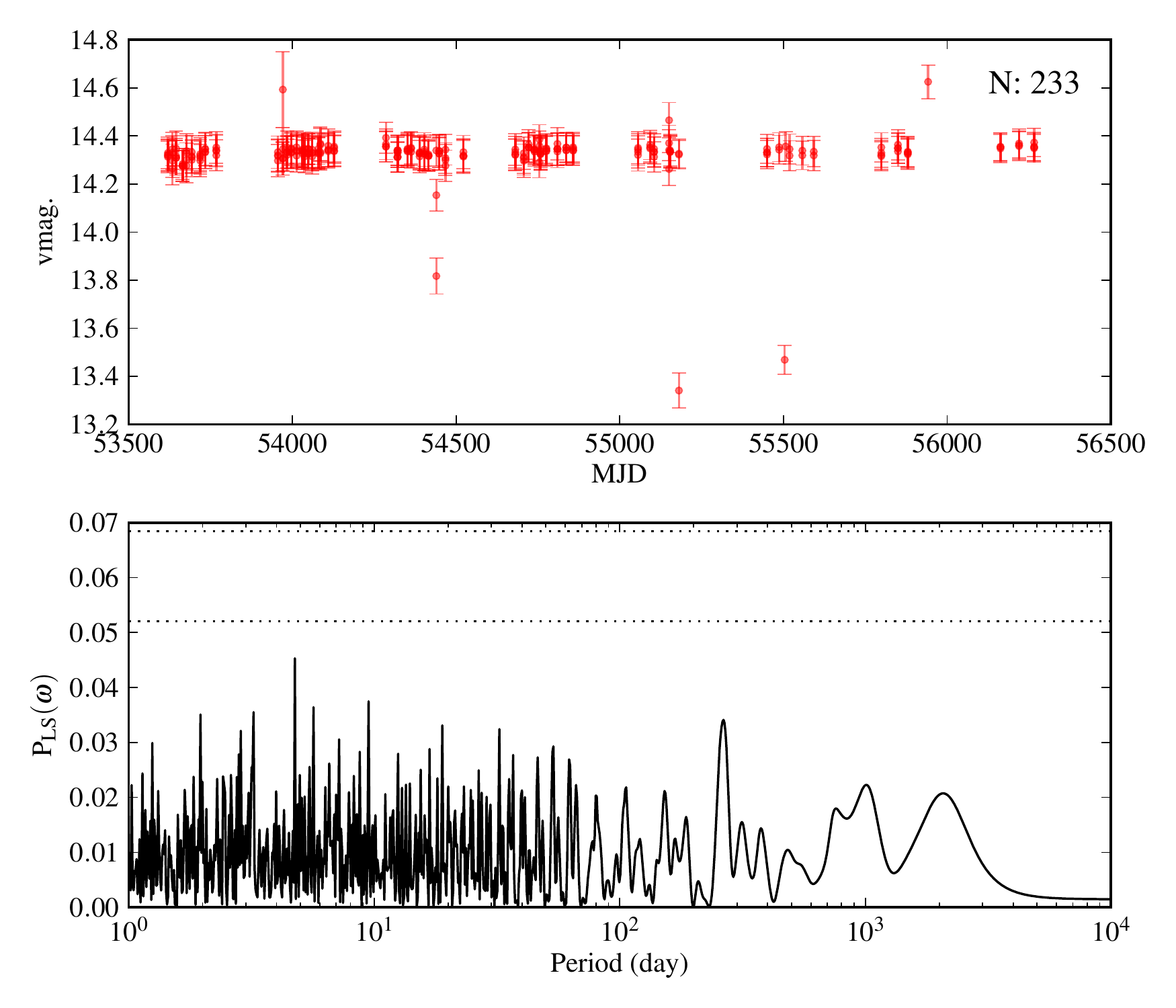}
\vspace{-0.2cm}
\caption{{\it Top panel:}  light curve of optical magnitudes for a candidate member in our sample (2MASS J02103888-4557248); N is the total number of data points.  {\it Bottom panel:} Lomb-Scargle periodogram for the data presented in the top panel. The dashed lines represent the levels of significance: 1\,$\sigma$ and 5\,$\sigma$.}
\label{fig:catalina_example}
\end{center}
\end{figure}

We used the Catalina survey \citep{Drake2009}, aimed at identifying optical transients, covering 33,000\,deg$^2$ of the sky, to analyse optical photometric variability.  We cross-matched our sample with the second data release via the online multi-object search service\footnote{\url{http://nesssi.cacr.caltech.edu/cgi-bin/getmulticonedb_release2.cgi}}. From this cross-match there were 235 matches.  This was significantly reduced to 55 matches, considering objects fainter than 13\,mag. to avoid saturation.  We checked the light curves of all 55 of these objects for any signs of variability using a Lomb-Scargle periodogram via the {\sc astroML} python package \citep{Vanderplas2012} and derived $V$ magnitudes, based on the median value of those data, for objects with 3 or more epochs of data (see Figure~\ref{fig:catalina_example} for an example).  However, the median values we derived were, in most cases, much higher (0.5-1.0\,mag.) than those catalogued in NOMAD and URAT1.   
This discrepancy is most likely down to the transformation from unfiltered photometry to $V$ magnitudes.  In this analysis we do not use the median magnitudes derived from Catalina photometry because of this large discrepancy.  However, any variability or periodicity seen in an objects light curve is included as a flag in Table~\ref{tab:online_appendix_tab} and is discussed further in Section~\ref{subsec:vmag_disc}. 

{
\begin{table}
\tiny
\caption{Surveys and catalogues cross-matched with the sample presented.}
\begin{tabular}{p{1.0cm} p{3.6cm} p{3.2cm}}
\hline\hline\\
  \multicolumn{1}{l}{Code} &
  \multicolumn{1}{l}{Reference} &
  \multicolumn{1}{l}{Parameters} \\[1ex]
\hline\\

NO04    &    \cite{Nordstrom2004}     &  RV \\
BA12    &    \cite{Bailey2012}        & RV  \\
EL14    &    \cite{Elliott2014}       &  RV   \\
MA14    &    \cite{Malo2014}          &  RV, H$_\alpha$, Li, L$_\mathrm{x}$  \\
RAVE    &    \cite{Kordopatis2013}    & RV  \\
DE15    &    \cite{Desidera2015}      & RV, Li\\
RO13    &    \cite{Rodriguez2013}     &  RV, H$_\alpha$, Li, NaI  \\
RI06    &    \cite{Riaz2006}     & RV,H$_\alpha$,L$_\mathrm{x}$/L$_\mathrm{bol}$,CaH1,2,3     \\
BE15 & \cite{Bell2015} & $V$ mag. \\
IPHAS & \cite{Barentsen2014} & H$_\alpha$\\
GALEX & \cite{Bianchi2011} & NUV, FUV mag.\\
URAT1 & \cite{Zacharias2015} & $R$,$B$,$V$ mag.\\
CATA & \cite{Drake2009} & $V$ mag. (+ variability)\\
ASAS & \cite{Pojmanski2002} & $V$ mag. (+ variability)\\
RO13 & \cite{Rodriguez2013} & RV \\
SH12 & \cite{Shkolnik2012} & RV\\
SC12 & \cite{Schlieder2012} & RV, SpT \\
SPM4 & \cite{Girard2011} & $B$,$V$ mag. \\
RE09 & \cite{Reiners2009} & RV, Li presence \\
RI15 & \cite{Riviere-Marichalar2015} & RV, H$_\alpha$ \\ 
M013 & \cite{Moor2013}  & RV, Li, H$_\alpha$,L$_\mathrm{x}$/L$_\mathrm{bol}$ \\
GO06 & \cite{Gontcharov2006} & RV, $V$ mag. \\
DE12 & \cite{deBruijne2012} & RV \\
FR13 & \cite{Frith2013} & RV \\
MO08 & \cite{Morin2008} & RV \\
LO10 & \cite{Looper2010} & RV \\
KH07 & \cite{Kharchenko2007} & RV \\
HO91 & \cite{Hoffleit1991} & RV \\
LE13 & \cite{Lepine2013} & H$_\alpha$, CaH2,3 \\
MO01a & \cite{Montes2001} & RV \\
LO06 & \cite{Lopez2006} & RV, Li \\
MA10 & \cite{Maldonado2010} & RV \\
WISE & \cite{Wright2010} & $W1$, $W2$, $W3$, $W4$ \\
RI14 & \cite{Riedel2014} & $V$ mag. \\
VI56 & \cite{Vyssotsky1956} & $V$ mag. \\
ME06 & \cite{Mermilliod2006} & $V$ mag. \\
LA01 & \cite{Lawson2001} & $V$ mag. \\
KH09 & \cite{Kharchenko2009} & $V$ mag. \\
NOMAD & \cite{Zacharias2005} & $B$,$V$,\,$R$ mag. \\
WDS  & \cite{Mason2001} & Multiplicity \\
MA13 & \cite{Malo2013} & RV,H$_\alpha$,Li \\
GSC & \cite{Lasker2008} & $B$,$U$,$V$ mag. \\
TYC & \cite{Hog2000} & $B$,$V$ mag. \\
3XMM & \cite{XMM2013} & X-ray: flux \\
CHAN & \cite{Evans2010S} & X-ray: flux \\
BSC & \cite{Voges1999} & X-ray: counts, HR \\
FSC & \cite{Voges2000} & X-ray: counts, HR \\
\hline
\end{tabular}
\label{tab:prev_surveys}
\end{table}}

\subsection{Previous searches for members of young associations}

Some of the works listed in Table~\ref{tab:prev_surveys} and additionally other works are specifically concerned with identifying new members of the young associations.  Below we note any overlap between our identified objects and other works.  We are interested in not only confirming the youth of the objects, but also in  analysing their radial velocity agreement with their associated known member.

\begin{itemize}
\item \cite{Malo2013}:  There were two objects identified here that are in common (2MASS J01034210+4051158, 2MASS J21100535-1919573). Both objects have compatible radial velocity values with their associated primary stars.  

\item \cite{Gagne2015}:  We found one object in common (2MASS J11020983-3430355), which was also identified in \cite{Teixeira2008} and is discussed in more detail in Appendix~\ref{sec:pre_identified_wide_comps}.  This object has compatible radial velocity with its associated primary.

\item \cite{Shkolnik2012}:  There were thre objects in common (2MASS J01034210+4051158, 2MASS J04373746-0229282, and 2MASS J12573935+3513194).  The radial velocity values were consistent with their primary stars for the first two sources.  The third source is a spectroscopic binary and therefore its radial velocity is not consistent with its associated primary star, see Appendix~\ref{sec:notes_individ_sources}.  Owing to a lack of data its system velocity cannot be derived at this time.

\item \cite{Schlieder2012}:  We found one object in common (2MASS J05015665+0108429), which has compatible radial velocity with its associated primary.

\item \cite{Rodriguez2013}: We found one object in common (2MASS J02105345-4603513), which is most likely a spectroscopic binary.  The target has an incompatible radial velocity and is discussed in more detail in Appendix~\ref{sec:notes_individ_sources}.

\item \cite{Kraus2014}: Although we included the bonafide member list from this work, we did not include identified candidate members.  There was one successful match with these candidate members (2MASS J00302572-6236015).  This source has compatible radial velocity with its associated primary. 

\item \cite{Alonso-Floriano2015}: The authors identified 36 potential wide binaries, and 16 of these systems only had one stellar component previously classified as a member of the $\beta$-Pic moving group. We recovered 15/16 of these new identifications.

 The one detection we did not recover we conclude is not a physical wide companion (V4046 Sgr -- GSC 07396-00759); see Appendix~\ref{sec:pre_identified_wide_comps} for details.  
\end{itemize}

\section{Spectroscopic parameters and observations}
\label{sec:new_spec}

High-resolution optical spectra allow us to compute radial velocity values and look for signatures of youth.  In the following analysis, we computed EWs for the gravity-sensitive atomic alkali lines of K\,I (7699\,\AA) and NaI (8194.8\,\AA), see \cite{Martin1997}. Additionally,  EWs of H$_\alpha$, associated with activity, and lithium which is used as an age indicator; see \cite{Barrado1999, Binks2014}. 

To produce the spectroscopic quantities described above we used the line-fitting method described in \cite{Bayo2011}.  This automatic line measurement procedure has been improved and includes a local pseudo-continuum iterative identification process and Monte Carlo solutions on this identification process to take  the normalisation contribution to the uncertainties of the line measurements into account.

\subsection{Existing observations}


We searched the public archives of four high-resolution spectrographs (UVES/VLT, FEROS/2.2m, HARPS/3.6m, and HIRES/KECK).   We found one HARPS spectrum for target 2MASS J23485048-2807157 (Prog. ID: 92.C-0224) and one HIRES spectrum for target 2MASS J02455260+0529240 (Prog. ID: U033Hr). 
The radial velocities were computed using a binary mask (SpT: K0) to produce a cross-correlation function (CCF) as described in \cite{Queloz1995, Elliott2014}. The spectroscopic quantities are shown in Table~\ref{tab:hermes_feros_obs}. We could not compute EWs for Na\,I or K\,I for the HARPS observation because of the wavelength coverage of the spectrum (3780--6910\,\AA).

\subsection{New observations}

To further classify identified candidates we obtained new high-resolution spectra using HERMES/Mercator and FEROS/2.2\,m.  The details of the observations are presented below.

{\it HERMES}: Spectroscopic observations were obtained at the 1.2\,m Mercator Telescope at the \emph{ Observatorio del Roque de los Muchachos}  (La Palma, Spain) on 14-18 December 2015 with HERMES, High Efficiency and Resolution Mercator Echelle Spectrograph \citep{Raskin2011}.
Using the high-resolution mode (HRF), the spectral resolution is 85000 and the wavelength range is from $\lambda$3800~{\AA} to $\lambda$8750~{\AA} approximately. All the obtained \emph{echelle} spectra were reduced with the automatic pipeline \citep{Raskin2011} for HERMES. We were limited to brighter targets to keep the integration times below 2000\,seconds for a signal-to-noise ratio (S/N) $\geq$30 at 6500\,\AA.  When the observations were taken we were not aware that 3/4 of these targets had existing radial velocity values.  However, with the newly obtained spectra we were able to compare our values to those of the literature, increase our sensitivity to single-lined spectroscopic binaries, and additionally calculate EWs of spectral lines.

{\it FEROS}:  We observed five newly identified candidate members using FEROS on 25-12-2015.  These observations were taken with the standard FEROS setup (2\arcsec fibre aperture, R$\sim$48,000, 3600-9200\,\AA) in {\sc objcal} mode, using the ThAr+Ne lamp for simultaneous wavelength calibration.  The targets were faint (V$\approx$13-14.8 mag.) and therefore the S/N was low ($\leq$20).  To calculate accurate radial velocities for these spectra, we used a dedicated pipeline built from a modular code (CERES, Brahm et al. in prep.) that is capable of processing data in
a homogeneous and optimal way coming from different
echelle spectrographs. A description of a similar pipeline
developed for the Coralie spectrograph can be found in \cite{Jordan2014}.  This FEROS pipeline is able to achieve a long-term RV precision of 5\,m/s in the high S/N limit. In the case of low S/N data ($\sim$20), the optimal extraction routine of the pipleline coupled with the careful treatment of systematics, allows it to obtain RV precisions of $\sim$20-30\,m/s. In order to increase the accuracy of the wavelength solution,
the bluest ($\lambda < 3900$\,\AA)and reddest ($\lambda > 6700$\,\AA) orders are not used. Some features such as the NaI and K\,I doublets are in the reddest orders and therefore are unavailable in these 1D reduced spectra. For the calculation of all EWs and indices from molecular bands, we used the standard reduced products from the MIDAS pipeline. Unfortunately the low S/N of our observations prevented us from continuum identification and extraction in several wavelengths regions. One of these regions was Li\,I (6708\,\AA), which is affected by a strong Ar emission line originated from the contiguous comparison fibre. We therefore do not analyse the Li\,I region from these FEROS spectra.

A summary of the observations and derived spectroscopic properties are shown in Table~\ref{tab:hermes_feros_obs} for the HERMES and FEROS observations.



{
\centering
\begin{table*}
{\tiny
\caption{Summary of targets and their derived properties for high-resolution spectra analysed in this work.}
\label{tab:hermes_feros_obs}
\begin{tabular}{p{2.3cm} p{1.9cm}p{1.6cm}p{1.6cm}p{1.6cm}p{0.9cm}p{0.8cm}p{0.7cm}p{0.6cm} p{1.0cm}}
\hline\\[-2ex]
\hline\\
  \multicolumn{1}{l}{2MASS ID} &
  \multicolumn{1}{l}{RV} &      
  \multicolumn{1}{l}{Na\,I (8194.8)} &  
  \multicolumn{1}{l}{K\,I (7699)} &   
  \multicolumn{1}{l}{H$_\alpha$} &
  \multicolumn{1}{l}{Li\,I} &  
  \multicolumn{1}{l}{T$_\mathrm{eff}$} &        
  \multicolumn{1}{l}{SpT} &
  \multicolumn{1}{l}{SpT ref.} &
  \multicolumn{1}{l}{Obs.}   \\
    \multicolumn{1}{l}{} &
  \multicolumn{1}{l}{(km\,s$^{-1}$)} &      
  \multicolumn{1}{l}{(\AA)} &
  \multicolumn{1}{l}{(\AA)} &   
  \multicolumn{1}{l}{(\AA)} &
  \multicolumn{1}{l}{(m\AA)} &  
  \multicolumn{1}{l}{(K)} &      
  \multicolumn{1}{l}{} &
  \multicolumn{1}{l}{} &
  \multicolumn{1}{l}{}   \\
\hline\\  
01334282-0701311 & -14.934$\pm$0.003 & 0.26$\pm$0.01 & 0.17$\pm$0.01  & 2.00$\pm$0.11 &  64$\pm$3 & 5830 & G1 & 1 & HERMES\\
01373545-0645375 & 11.689$\pm$0.002 & 0.329$\pm$0.026 & 0.210$\pm$0.003 & 1.202$\pm$0.101 &  101$\pm$10 & 5200 & G6 & 2 & HERMES\\
03281095+0409075 & 10.952$\pm$0.667 & 0.146$\pm$0.048  & 0.196$\pm$0.044 & 2.727$\pm$0.329 & 124$\pm$16 & 5900 & F6 & 2 & HERMES\\
04373746-0229282 & 22.047$\pm$0.031   & 1.052$\pm$1.162 & 0.769$\pm$0.034 & -2.142$\pm$0.127 & 61$\pm$23 & 3750 & M1.1  & 3 & HERMES\\
01354915-0753470 & 6.487$\pm$0.029 &  0.69$\pm$0.09 & 0.69$\pm$0.05 & -0.06$\pm$0.02 & \ldots & 3500 & M4 & 4  & FEROS\\
03074909-2750467 & 16.512$\pm$0.171 & 1.21$\pm$0.08 & 1.21$\pm$0.13 & -12.80$\pm$1.11 & \ldots & 3500  & M4 & 4  & FEROS\\
04515223-4640497 & 51.188$\pm$0.019  &  0.35$\pm$0.03 & 0.49$\pm$0.05 & -0.03$\pm$0.01 & \ldots & 4800  & K6 & 4   & FEROS\\
07013945-4231370 & 24.339$\pm$0.025  &  -0.001$\pm$0.002 & 0.45$\pm$0.07  & 0.33$\pm$0.22 & \ldots & 4300  & K4  & 4  & FEROS\\
08371456-5517343 & 81.314$\pm$0.019 & -0.15$\pm$0.07  &   0.001$\pm$0.001 & 0.70$\pm$0.10 & \ldots & 4850  & K9 & 4  & FEROS\\
23485048-2807157 & 7.430$\pm$0.032  & \ldots & \ldots & 1.63$\pm$0.17 & 140$\pm$10 & 5200  & G8  & 2 & HARPS\\
02455260+0529240 & 88.502$\pm$0.053  & \ldots & 0.912$\pm$0.039 & -2.462$\pm$0.216 & \ldots & 3100  & M1 & 2 & HIRES\\
\hline\\[-0.1ex]
\end{tabular}}
{\footnotesize \bf References.}{\footnotesize~1: \cite{Gray2006}, 2: T$_\mathrm{eff}$ conversion from VOSA spectral energy distribution \citep{Bayo2008}, 3: \cite{Shkolnik2012}, 4: estimation from FEROS data reduction.}\label{tab:}
\end{table*}}

\section{Properties of candidates from our search}
\label{sec:cand_properties}

Discussions of individual objects with discrepant/dubious spectroscopic parameters (denoted by a "N" or "?" in {\it Member?} column of Table~\ref{tab:identified_comps}) can be found in Appendix~\ref{sec:notes_individ_sources}. The detailed spectroscopic, kinematic and photometric properties of all identified targets and their respective references can be found in Table~\ref{tab:online_appendix_tab}. For a brief summary of the 84 targets identified in this work and their membership status see Table~\ref{tab:identified_comps}.  Below we compare individual kinematic, spectroscopic, and photometric properties of identified targets with our original sample and other young sources to assess the success of our method.

Spectral types were calculated by converting effective temperatures using the relationship for pre-main-sequence stars from \cite{Pecaut2013}.  The effective temperatures were derived from fitting the SEDs (produced using the VOSA tool; \citealt{Bayo2008}) with the evolutionary models of \cite{Baraffe2015}.

\subsection{Identification of existing members as wide companions to other known members}

Using our method we identified known members of the moving groups as potential wide companions to other known members in our sample.  

\begin{figure}
\begin{center}
\includegraphics[width=0.49\textwidth]{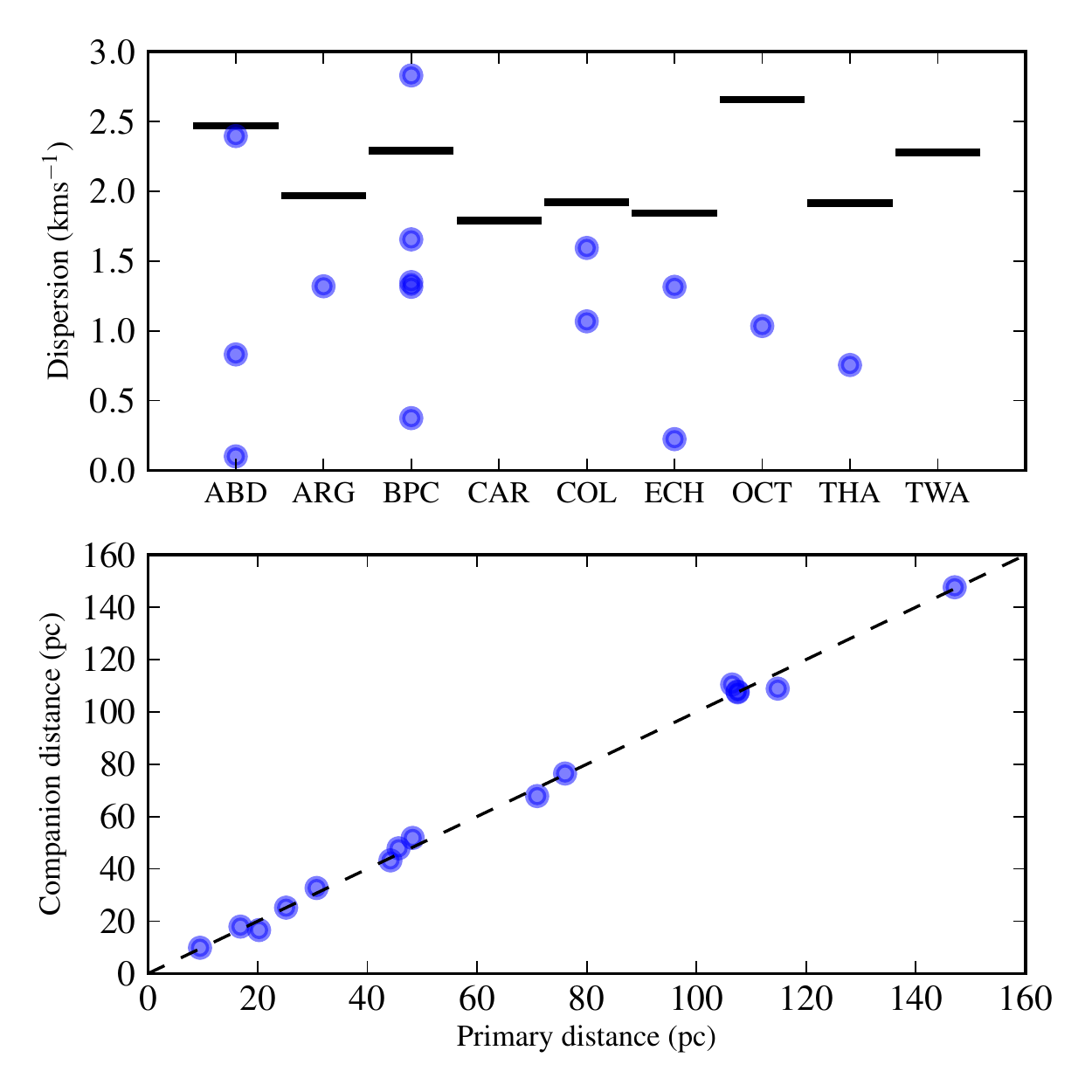}
\vspace{-0.2cm}
\caption{{\it Top panel}: black markers represent the magnitude of the standard deviation in Galactic velocities (U, V, W) for each association.  The blue markers are the absolute difference between companions and their associated primary star, as shown in Table~\ref{tab:prev_identified_members}. {\it Bottom panel}: photometric distances derived from the convergence method for companions and their associated known member.}
\label{fig:galac_disp_prev_pairs}
\end{center}
\end{figure}

This effectively reduced our sample size of primaries from 581 to 567.  We used these sources to check whether our method based on 2MASS photometry and proper motion is consistent for sources with derived photometric distances and Galactic velocities. Table~\ref{tab:prev_identified_members} summarises the results and notes any companions to the sources that could affect their photometry ($<$3\arcsec) and therefore their distance estimation.  The uncertainties on the Galactic velocities are $\sim$1\,kms$^{-1}$.  To quantify the agreement between components in these 14 cases, we first calculated the standard deviation of (U, V, W) Galactic velocities for the known members in each association. We compared this to the magnitude of the differences between the Galactic velocities grouped as in Table~\ref{tab:prev_identified_members}.  We found that 13/14 targets had dispersions below the standard deviation of their respective association, the left panel of Figure~\ref{fig:galac_disp_prev_pairs} shows this result graphically.  Additionally, the right panel shows the photometric distances derived between the primary and its associated companion for all 14 systems.  These two results are a very good indication that these targets do in fact form physically wide multiple systems rather than being the result of a projection effect due to the spatial density of stars.  The formation, abundance, and dynamical evolution of wide binaries in the young associations will be discussed in further work (Elliott et al. in prep.).  We do not refer to these targets as {\it identified in this work}, they form part of the {\it original sample} in any discussion presented here.

\subsection{Radial velocities}

\begin{figure}[h]
\begin{center}
\includegraphics[width=0.49\textwidth]{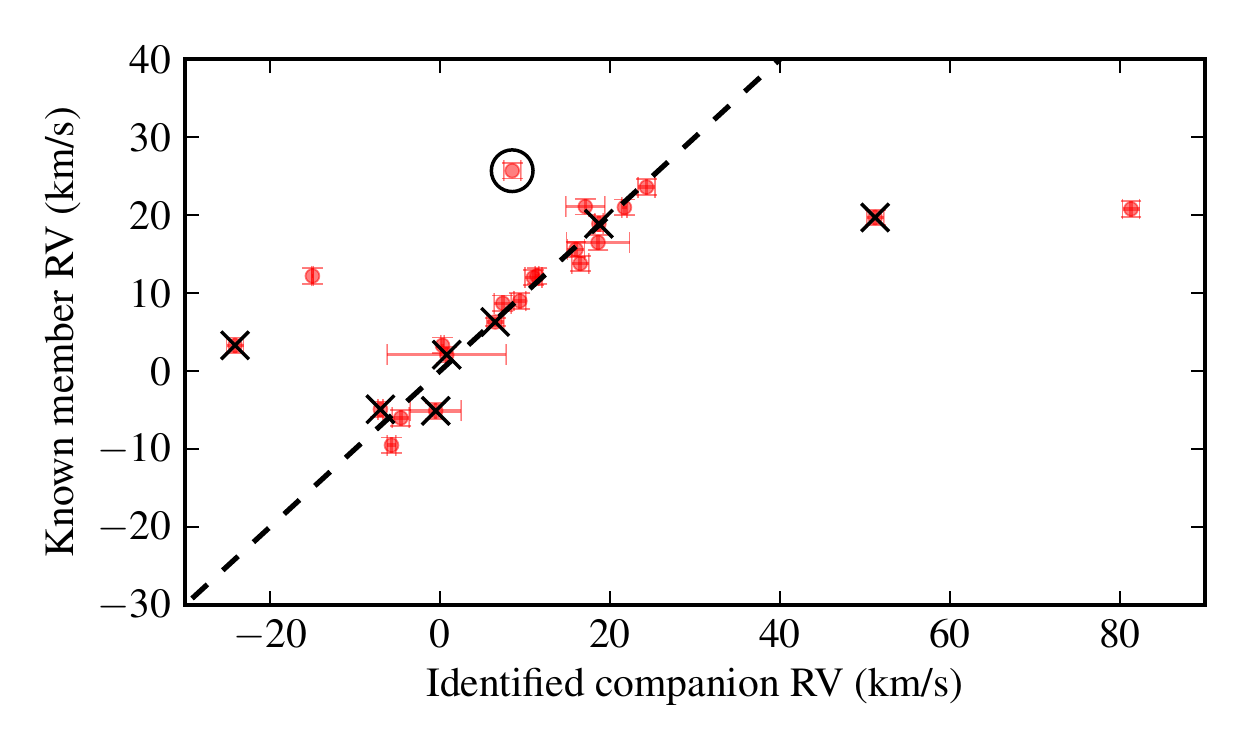}
\vspace{-0.2cm}
\caption{Radial velocity of identified targets in this work versus radial velocity our their associated known member.  The dotted line represents a 1:1 relation in the quantities.  Crosses indicate the known member is a spectroscopic binary (plotted radial velocity is the system(gamma) velocity). The open circle represents the suspected single-lined spectroscopic binary 2MASS J02105345-4603513.}
\label{fig:rv_comp}
\end{center}
\end{figure}

In Figure~\ref{fig:rv_comp} we show a comparison of radial velocities between targets identified in this work (companions) and their associated known member for the 23 targets with both measurements.  If these companions formed very wide binary systems with their associated known member, we would expect a 1:1 relation between the two quantities within uncertainties.  As can be seen, we do indeed see this relation for the majority of targets (18/23).  As these targets share the same proper motion and radial velocity, by assuming the same distance, they share the same Galactic velocity.  Some of the known members are part of spectroscopic binary systems, these are indicated by crosses in Figure~\ref{fig:rv_comp}.  The plotted radial velocities of the known members are the gamma (system) velocities and therefore the effect of the spectroscopic component has been accounted for.  The quality of this correction depends on how much of the orbital phase has been covered by previous radial velocity observations.  However, as shown by the agreement between the values, this does not have a significant effect.  The suspected single-lined spectroscopic binary 2MASS J02105345-4603513 is highlighted by an open black circle and discussed, along with the other four discrepant targets in Appendix~\ref{sec:notes_individ_sources}.  

\subsection{Identification of young objects combining GALEX, 2MASS, and WISE photometry}
\label{sec:uv_criteria}

Some previous works have focussed on identifying new late-type members (M0-M5) detecting their predicted enhanced UV emission \citep{Shkolnik2012, Rodriguez2013}. The convective envelopes of these low-mass stars stars combined with their differential rotation produces strong magnetic dynamos. These dynamos lead to enhanced chromospheric and coronal activity, which increases UV emission and X-ray emission (see Section~\ref{sec:x_ray}), respectively.

{\tiny
\onecolumn
\LTcapwidth=\textwidth
\begin{longtable}{p{2.3cm} p{0.8cm}p{0.6cm}p{0.6cm}p{0.6cm}p{0.6cm} p{1.4cm} p{1.5cm} p{0.5cm} p{2.7cm}p{1cm}p{1.0cm}}
\caption{Basic properties and membership status of targets identified in this work.}\label{tab:identified_comps}\\
\hline\\
  \multicolumn{1}{l}{2MASS identifier} &
  \multicolumn{1}{l}{Dist.\,\tablefoottext{a}} &      
  \multicolumn{1}{l}{SpT} &  
  \multicolumn{1}{l}{$V$} &    
  \multicolumn{1}{l}{$K$} &
  \multicolumn{1}{l}{Mass} &  
  \multicolumn{1}{l}{$\mu_{\alpha, *}$}  &
  \multicolumn{1}{l}{$\mu_\delta$}   &  
  \multicolumn{1}{l}{Ass.}  &
  \multicolumn{1}{l}{Quantity\,\tablefoottext{b}}  &
  \multicolumn{1}{l}{Ref.}  &    
  \multicolumn{1}{l}{Mem?\,\tablefoottext{c}}  \\
  \multicolumn{1}{l}{} &
  \multicolumn{1}{l}{(pc)} &      
  \multicolumn{1}{l}{} &  
  \multicolumn{1}{l}{(mag.)} &    
  \multicolumn{1}{l}{(mag.)} &
  \multicolumn{1}{l}{(M$_\odot$)} &  
  \multicolumn{1}{l}{(mas/yr)} &  
  \multicolumn{1}{l}{(mas/yr)} &
  \multicolumn{1}{l}{} &
  \multicolumn{1}{l}{}   &
  \multicolumn{1}{l}{}  &
  \multicolumn{1}{l}{}   \\
\hline\\
\endfirsthead
\hline\hline\\
  \multicolumn{1}{l}{2MASS identifier} &
  \multicolumn{1}{l}{Dist.\,\tablefoottext{a}} &      
  \multicolumn{1}{l}{SpT} &  
  \multicolumn{1}{l}{$V$} &    
  \multicolumn{1}{l}{$K$} &
  \multicolumn{1}{l}{Mass} &  
  \multicolumn{1}{l}{$\mu_{\alpha, *}$}  &
  \multicolumn{1}{l}{$\mu_\delta$}   &  
  \multicolumn{1}{l}{Ass.}  &
  \multicolumn{1}{l}{Quantity\,\tablefoottext{b}}  &
  \multicolumn{1}{l}{Ref.}  &    
  \multicolumn{1}{l}{Mem?\,\tablefoottext{c}}  \\
  \multicolumn{1}{l}{} &
  \multicolumn{1}{l}{(pc)} &      
  \multicolumn{1}{l}{} &  
  \multicolumn{1}{l}{(mag.)} &    
  \multicolumn{1}{l}{(mag.)} &
  \multicolumn{1}{l}{(M$_\odot$)} &  
  \multicolumn{1}{l}{(mas/yr)} &  
  \multicolumn{1}{l}{(mas/yr)} &
  \multicolumn{1}{l}{} &
  \multicolumn{1}{l}{}   &
  \multicolumn{1}{l}{}  &
  \multicolumn{1}{l}{}   \\
\hline\\
\endhead
00341993+2528147 & 49.9  &  M4  &               \ldots  &  10.7  &  0.23 &   90.7\,$\pm$\,4.1  &  -105.3\,$\pm$\,4.1 &  ABD  &    &    &  \ldots\\
01034210+4051158 & 29.6  &  K7  &               12.9  &  8.5  &  0.41 &   132.0\,$\pm$\,8.0  &  -164.0\,$\pm$\,8.0 &  ABD  &  H$_\alpha$, X, CaH, UV  &  1, 5  &  Y\\
02103888-4557248 & 73.7  &  K5  &               14.6  &  11.4  &  0.25 &   48.2\,$\pm$\,1.6  &  -1.5\,$\pm$\,1.6 &  ABD  &  Fl  &  5  &  \ldots\\
02105345-4603513 & 73.7  &  M4  &               15.2  &  10.3  &  0.44 &   53.2\,$\pm$\,1.8  &  -10.2\,$\pm$\,1.8 &  ABD  &  RV, X, UV  &  2  &  Y?\\
02123342+2356379 & 36.6  &  M4  &               17.0  &  11.0  &  0.12 &   131.0\,$\pm$\,8.0  &  -158.0\,$\pm$\,8.0 &  ABD  &    &    &  \ldots\\
02484213+2716185 & 50.7  &  M5  &               18.8  &  13.8  &  0.04 &   62.4\,$\pm$\,5.2  &  -127.3\,$\pm$\,5.2 &  ABD  &    &    &  \ldots\\
03331403+4615194 & 34.3  &  F5  &               11.8  &  7.6  &  0.63 &   64.7\,$\pm$\,4.0  &  -172.4\,$\pm$\,3.6 &  ABD  &  RV, H$_\alpha$, CaH  &  14  &  Y\\
04080429+0143548 & 55.3  &  K9  &               14.5  &  10.9  &  0.23 &   29.9\,$\pm$\,4.6  &  -88.7\,$\pm$\,5.2 &  ABD  &    &    &  \ldots\\
04575034-0652070 & 45.0  &  M2  &               15.6  &  11.3  &  0.14 &   41.3\,$\pm$\,4.3  &  -103.1\,$\pm$\,5.2 &  ABD  &    &    &  \ldots\\
05060187-1543564 & 78.9  &  M1  &               \ldots  &  14.1  &  0.05 &   18.7\,$\pm$\,4.0  &  -50.2\,$\pm$\,4.1 &  ABD  &    &    &  \ldots\\
05105505-0410563 & 76.9  &  M4  &               15.9  &  10.8  &  0.38 &   19.0\,$\pm$\,3.4  &  -54.4\,$\pm$\,3.7 &  ABD  &    &    &  \ldots\\
05274075-0930438 & 21.9  &  M3  &               \ldots  &  9.7  &  0.14 &   61.6\,$\pm$\,5.1  &  -172.7\,$\pm$\,5.1 &  ABD  &    &    &  \ldots\\
05454276-1450198 & 75.8  &  M4  &               \ldots  &  13.6  &  0.06 &   10.7\,$\pm$\,4.1  &  -51.6\,$\pm$\,4.1 &  ABD  &    &    &  \ldots\\
07285117-3015527 & 15.7  &  M4  &               13.4  &  8.1  &  0.25 &   -45.7\,$\pm$\,3.4  &  17.2\,$\pm$\,3.5 &  ABD  &  RV  &  9  &  Y\\
07435039-7940252 & 80.6  &  K9  &               16.2  &  12.4  &  0.15 &   -6.7\,$\pm$\,3.8  &  49.3\,$\pm$\,3.9 &  ABD  &    &    &  \ldots\\
07453522-7940136 & 80.6  &  F5  &               12.1  &  8.9  &  0.82 &   -16.5\,$\pm$\,2.5  &  42.2\,$\pm$\,2.5 &  ABD  &    &    &  \ldots\\
09281506-7815223 & 116.5  &  F5  &              8.6  &  7.7  &  0.22 &   -25.8\,$\pm$\,1.1  &  21.9\,$\pm$\,1.1 &  ABD  &    &    &  \ldots\\
12573935+3513194 & 19.3  &  M4  &               13.0  &  8.0  &  0.33 &   -281.8\,$\pm$\,3.6  &  -147.0\,$\pm$\,6.7 &  ABD  &  RV, CaH, H$_\alpha$  &  4, 14  &  Y?\\
17360095-1320580 & 46.3  &  M0  &               15.3  &  11.7  &  0.11 &   -4.1\,$\pm$\,6.0  &  -108.7\,$\pm$\,6.3 &  ABD  &    &    &  \ldots\\
23270114+0055200 & 47.2  &  K5  &               14.0  &  10.6  &  0.23 &   80.3\,$\pm$\,3.9  &  -99.2\,$\pm$\,3.9 &  ABD  &    &    &  \ldots\\
23485048-2807157 & 42.2  &  G8  &               9.1  &  7.1  &  0.87 &   99.2\,$\pm$\,0.9  &  -100.9\,$\pm$\,1.7 &  ABD  &  RV, Li, H$_\alpha$, X  &  5  &  Y\\
23514340+3127045 & 42.8  &  M4  &               14.1  &  9.5  &  0.38 &   96.5\,$\pm$\,3.3  &  -86.6\,$\pm$\,3.6 &  ABD  &  UV  &  5  &  Y\\
\hline\\[0.2ex]
03262220+2313122 & 16.9  &  M4  &               16.4  &  10.7  &  0.04 &   247.0\,$\pm$\,8.0  &  -96.0\,$\pm$\,8.0 &  ARG  &  Fl  &  5  &  \ldots\\
06494841-2858025 & 120.8  &  M4  &              15.6  &  11.5  &  0.30 &   -15.7\,$\pm$\,4.3  &  20.0\,$\pm$\,4.3 &  ARG  &    &    &  \ldots\\
07013945-4231370 & 101.0  &  K4  &              13.9  &  10.9  &  0.38 &   -9.9\,$\pm$\,1.1  &  34.6\,$\pm$\,1.1 &  ARG  &  KI, RV, H$_\alpha$  &  5  &  Y\\07283006-4908589 & 82.8  &  F8  &             8.8  &  7.6  &  1.18 &   -25.4\,$\pm$\,0.8  &  48.8\,$\pm$\,1.3 &  ARG  &  RV, Li  &  5, 6  &  Y\\
08374680-5252307 & 152.3  &  K2  &              14.5  &  12.3  &  0.28 &   1.7\,$\pm$\,1.9  &  6.2\,$\pm$\,1.8 &  ARG  &    &    &  \ldots\\
09394669-2133023 & 93.3  &  M5  &               18.0  &  14.4  &  0.04 &   -41.4\,$\pm$\,4.0  &  15.8\,$\pm$\,4.0 &  ARG  &    &    &  \ldots\\
\hline\\[0.2ex]
01334282-0701311 & 24.0  &  G1  &               5.8  &  4.3  &  0.97 &   175.2\,$\pm$\,1.0  &  -82.3\,$\pm$\,1.0 &  BPC  &  RV, Li  &  5, 16  &  N\\
01354915-0753470 & 37.9  &  M4  &               13.8  &  9.8  &  0.13 &   88.6\,$\pm$\,5.1  &  -42.2\,$\pm$\,5.1 &  BPC  &  H$_\alpha$, KI, RV  &  5  &  Y?\\01373545-0645375 & 24.0  &  G6  &            7.7  &  5.8  &  0.89 &   171.3\,$\pm$\,1.0  &  -98.5\,$\pm$\,1.0 &  BPC  &  X, RV, Li, H$_\alpha$  &  5, 15, 16  &  Y\\
02160734+2856530 & 39.5  &  M2  &               15.7  &  11.6  &  0.04 &   92.9\,$\pm$\,3.1  &  -64.0\,$\pm$\,3.2 &  BPC  &    &    &  \ldots\\
04373746-0229282 & 29.4  &  M1  &               10.6  &  6.4  &  0.78 &   45.9\,$\pm$\,1.3  &  -63.6\,$\pm$\,1.2 &  BPC  &  Li, RV, CaH, H$_\alpha$, KI, X  &  1, 4  &  Y\\
05015665+0108429 & 24.2  &  M4  &               12.9  &  7.7  &  0.28 &   33.2\,$\pm$\,2.2  &  -89.1\,$\pm$\,3.1 &  BPC  &  X, RV, CaH, H$_\alpha$  &  5, 7, 14  &  Y\\
05195327+0617258 & 71.0  &  M5  &               18.3  &  12.4  &  0.05 &   13.7\,$\pm$\,4.0  &  -38.1\,$\pm$\,4.0 &  BPC  &  X  &  5  &  Y\\
14141700-1521125 & 30.2  &  M3  &               15.6  &  8.8  &  0.18 &   -117.4\,$\pm$\,8.0  &  -196.6\,$\pm$\,8.0 &  BPC  &  UV  &  5  &  Y\\
17483374-5306118 & 71.0  &  M2  &               13.7  &  9.3  &  0.47 &   -2.5\,$\pm$\,2.0  &  -51.7\,$\pm$\,2.0 &  BPC  &    &    &  \ldots\\
18011138-5125594 & 48.1  &  M0  &               14.8  &  11.3  &  0.07 &   -9.8\,$\pm$\,5.2  &  -84.5\,$\pm$\,5.1 &  BPC  &    &    &  \ldots\\
18420483-5554126 & 51.9  &  M4  &               15.1  &  9.9  &  0.20 &   13.4\,$\pm$\,3.5  &  -73.9\,$\pm$\,3.7 &  BPC  &  H$_\alpha$, X, CaH, UV  &  1, 5  &  Y\\
18480637-6213470 & 52.4  &  F6  &               7.3  &  6.1  &  1.26 &   16.1\,$\pm$\,2.0  &  -80.3\,$\pm$\,2.0 &  BPC  &  Li, RV  &  8  &  Y\\
18580464-2953320 & 82.6  &  M3  &               12.8  &  8.8  &  0.75 &   15.4\,$\pm$\,2.4  &  -45.6\,$\pm$\,2.4 &  BPC  &  RV  &  8  &  Y\\
19223409-5429181 & 48.3  &  M0  &               14.0  &  10.7  &  0.10 &   32.0\,$\pm$\,2.4  &  -87.8\,$\pm$\,2.5 &  BPC  &    &    &  \ldots\\
20085122-2740536 & 48.0  &  M3  &               16.3  &  11.8  &  0.05 &   40.4\,$\pm$\,4.0  &  -61.7\,$\pm$\,4.0 &  BPC  &    &    &  \ldots\\
20321797-2600432 & 48.3  &  M5  &               18.6  &  13.0  &  0.01 &   50.0\,$\pm$\,4.9  &  -70.1\,$\pm$\,4.9 &  BPC  &    &    &  \ldots\\
21100461-1920302 & 32.6  &  M4  &               13.1  &  7.6  &  0.47 &   87.0\,$\pm$\,1.4  &  -94.4\,$\pm$\,3.1 &  BPC  &  RV, UV  &  5, 9  &  Y\\
21212446-6654573 & 30.2  &  G9  &               9.0  &  6.4  &  0.84 &   90.5\,$\pm$\,1.0  &  -90.9\,$\pm$\,1.6 &  BPC  &  X, RV  &  5, 17  &  Y?\\
\hline\\[0.2ex]
02590322-4232450 & 103.3  &  M3  &              15.0  &  10.3  &  0.52 &   41.1\,$\pm$\,2.3  &  0.0\,$\pm$\,1.5 &  CAR  &  UV  &  5  &  Y\\
03074909-2750467 & 54.6  &  M4  &               12.5  &  10.2  &  0.24 &   59.6\,$\pm$\,5.3  &  -16.4\,$\pm$\,17.6 &  CAR  &  RV  &  5  &  Y\\
08371456-5517343 & 180.8  &  K9  &              13.9  &  10.3  &  0.79 &   -7.9\,$\pm$\,1.8  &  9.7\,$\pm$\,1.9 &  CAR  &  H$_\alpha$, RV  &  5  &  N\\
09594633-7227360 & 83.6  &  M2  &               15.8  &  12.0  &  0.13 &   -37.4\,$\pm$\,3.4  &  23.0\,$\pm$\,3.4 &  CAR  &    &    &  \ldots\\
\hline\\[0.2ex]
01525509-5222266 & 83.4  &  M1  &               17.0  &  12.4  &  0.09 &   52.6\,$\pm$\,7.0  &  -7.5\,$\pm$\,7.0 &  COL  &    &    &  \ldots\\
02005969-1608428 & 80.9  &  M4  &               18.0  &  13.9  &  0.04 &   51.1\,$\pm$\,4.9  &  -25.0\,$\pm$\,4.9 &  COL  &    &    &  \ldots\\
02015050-1614575 & 80.9  &  M2  &               16.7  &  13.7  &  0.04 &   50.0\,$\pm$\,3.8  &  -30.6\,$\pm$\,3.8 &  COL  &    &    &  \ldots\\
03212872-1048588 & 128.1  &  M3  &              \ldots  &  13.7  &  0.08 &   24.7\,$\pm$\,4.0  &  -28.2\,$\pm$\,4.0 &  COL  &    &    &  \ldots\\
03255277-3601161 & 107.5  &  M4  &              15.4  &  10.6  &  0.47 &   33.9\,$\pm$\,2.8  &  -4.1\,$\pm$\,2.8 &  COL  &    &    &  \ldots\\
03281095+0409075 & 81.5  &  F6  &               8.8  &  7.2  &  1.26 &   46.7\,$\pm$\,0.7  &  -37.0\,$\pm$\,0.8 &  COL  &  RV, Li  &  5  &  Y\\
03400260+5441003 & 35.4  &  M5  &               17.7  &  12.0  &  0.04 &   102.0\,$\pm$\,4.0  &  -109.2\,$\pm$\,4.0 &  COL  &    &    &  \ldots\\
03471077+5138276 & 57.5  &  M2  &               15.6  &  11.8  &  0.09 &   60.1\,$\pm$\,3.7  &  -67.5\,$\pm$\,3.8 &  COL  &    &    &  \ldots\\
04423261+0905471 & 105.6  &  M3  &              15.8  &  11.0  &  0.37 &   31.7\,$\pm$\,3.9  &  -41.3\,$\pm$\,3.9 &  COL  &    &    &  \ldots\\
04515223-4640497 & 68.1  &  K6  &               14.3  &  11.7  &  0.13 &   27.7\,$\pm$\,2.5  &  17.1\,$\pm$\,2.6 &  COL  &  H$_\alpha$, RV  &  5  &  N\\
05005830-4100023 & 76.3  &  M4  &               14.7  &  9.6  &  0.54 &   32.6\,$\pm$\,1.5  &  9.6\,$\pm$\,2.8 &  COL  &    &    &  \ldots\\
05301939-1916161 & 115.1  &  K2  &              \ldots  &  9.1  &  0.86 &   17.8\,$\pm$\,3.1  &  -0.6\,$\pm$\,2.3 &  COL  &    &    &  \ldots\\
05374880+0233088 & 70.8  &  M3  &               14.2  &  9.5  &  0.52 &   16.8\,$\pm$\,3.7  &  -44.6\,$\pm$\,4.0 &  COL  &    &    &  \ldots\\
\hline\\[0.2ex]
09151348-7605020 & 117.5  &  K2  &              9.4  &  6.9  &  1.17 &   -31.2\,$\pm$\,1.0  &  23.3\,$\pm$\,1.0 &  ECH  &    &    &  \ldots\\
12201867-7417202 & 114.8  &  M0  &              15.0  &  11.6  &  0.07 &   -39.5\,$\pm$\,3.1  &  7.7\,$\pm$\,3.0 &  ECH  &    &    &  \ldots\\
\hline\\[0.2ex]
02174439+3330409 & 34.5  &  M3  &               16.3  &  11.8  &  0.05 &   44.8\,$\pm$\,4.1  &  -52.5\,$\pm$\,4.1 &  OCT  &    &    &  \ldots\\
06251922-6630435 & 126.1  &  M2  &              16.4  &  12.1  &  0.22 &   -13.0\,$\pm$\,3.8  &  25.4\,$\pm$\,3.3 &  OCT  &    &    &  \ldots\\
17420077-8608464 & 132.0  &  K5  &              12.9  &  9.6  &  0.80 &   9.2\,$\pm$\,1.2  &  -25.5\,$\pm$\,1.2 &  OCT  &    &    &  \ldots\\
23315208+1956142 & 6.1  &  M4  &                10.3  &  5.3  &  0.26 &   543.2\,$\pm$\,1.6  &  -44.6\,$\pm$\,1.6 &  OCT  &  H$_\alpha$, RV, CaH, X  &  5, 10, 14  &  Y\\
\hline\\[0.2ex]
00230961-6139143 & 44.6  &  M3  &               16.7  &  12.2  &  0.04 &   79.8\,$\pm$\,2.7  &  -49.3\,$\pm$\,2.6 &  THA  &    &    &  \ldots\\
00302572-6236015 & 42.5  &  M4  &               12.2  &  7.5  &  0.71 &   95.2\,$\pm$\,0.9  &  -48.0\,$\pm$\,0.9 &  THA  &  Li, H$_\alpha$, RV  &  11  &  Y\\02455260+0529240 & 54.3  &  M1  &            13.4  &  9.2  &  0.41 &   71.9\,$\pm$\,2.9  &  -41.5\,$\pm$\,3.0 &  THA  &  RV, UV  &  5  &  N\\
03153814-0339004 & 49.2  &  M3  &               16.8  &  12.1  &  0.05 &   65.4\,$\pm$\,4.2  &  -47.4\,$\pm$\,4.2 &  THA  &    &    &  \ldots\\
03365153-4957314 & 43.3  &  K5  &               12.4  &  9.8  &  0.20 &   124.5\,$\pm$\,2.2  &  -75.7\,$\pm$\,2.1 &  THA  &    &    &  \ldots\\
03484041-3738198 & 50.7  &  M2  &               13.3  &  8.7  &  0.51 &   79.0\,$\pm$\,1.1  &  -4.5\,$\pm$\,1.1 &  THA  &  X, RV  &  5, 12  &  Y\\
04354718-1210374 & 62.3  &  M3  &               \ldots  &  12.0  &  0.08 &   58.0\,$\pm$\,4.8  &  -26.4\,$\pm$\,4.2 &  THA  &    &    &  \ldots\\
04485254-5043145 & 54.8  &  M3  &               16.9  &  12.2  &  0.06 &   43.6\,$\pm$\,5.3  &  19.0\,$\pm$\,5.0 &  THA  &    &    &  \ldots\\
05154763-0931041 & 77.5  &  M5  &               17.8  &  14.0  &  0.03 &   39.5\,$\pm$\,5.6  &  -8.0\,$\pm$\,5.6 &  THA  &    &    &  \ldots\\
06350229-6951519 & 68.4  &  K2  &               11.9  &  8.6  &  0.72 &   21.8\,$\pm$\,1.5  &  41.6\,$\pm$\,1.4 &  THA  &  RV  &  9  &  Y\\
\hline\\[0.2ex]
11020983-3430355 & 51.0  &  M8  &               21.2  &  11.9  &  0.02 &   -77.0\,$\pm$\,9.5  &  -19.5\,$\pm$\,9.5 &  TWA  &  H$_\alpha$, CaH  &  13  &  Y\\
11130416-4516056 & 96.2  &  M0  &               15.7  &  11.7  &  0.08 &   -41.0\,$\pm$\,3.2  &  4.0\,$\pm$\,3.0 &  TWA  &  Fl  &  5  &  \ldots\\
12073145-3310222 & 53.8  &  K9  &               14.5  &  11.0  &  0.06 &   -62.7\,$\pm$\,1.8  &  -32.3\,$\pm$\,1.4 &  TWA  &  Fl  &  5  &  \ldots\\
12090628-3247453 & 53.8  &  M1  &               16.1  &  12.0  &  0.02 &   -68.1\,$\pm$\,2.0  &  -40.2\,$\pm$\,2.7 &  TWA  &    &    &  \ldots\\
12354893-3950245 & 73.0  &  M5  &               12.1  &  8.9  &  0.42 &   -48.6\,$\pm$\,1.7  &  -21.3\,$\pm$\,1.6 &  TWA  &  X  &  5  &  Y\\

\hline\\
\end{longtable}
}
\vspace{-0.5cm}
\noindent{\bf \footnotesize Notes. }{\footnotesize $^{(a)}$ Distance taken from the bonafide member that the target is associated with.}{\footnotesize $^{(b)}$ Optical flaring from Catalina photometry (Fl) is not used to assess the membership of targets.}{\footnotesize $^{(c)}$ Based on quantities listed in column {\it Quantity}: consistent radial velocity (RV), H$_\alpha$ EW (H$_\alpha$), Li absorption (Li), X-ray emission (X), UV/IR photometric criteria discussed in Section~\ref{sec:uv_criteria}: UV, CaH indices (CaH). Those systems with either a flag of "?" or "N" are discussed in Appendix~\ref{sec:notes_individ_sources}. \\}
\noindent{\bf \footnotesize References. }{\footnotesize 1: \cite{Riaz2006}, 2: \cite{Rodriguez2013}, 3: \cite{Kharchenko2007}, 4: \cite{Shkolnik2012}, 5:       This work, 6: \cite{Desidera2015}, 7: \cite{Schlieder2012}, 8: \cite{Moor2013}, 9: \cite{Kordopatis2013}, 10: \cite{Morin2008}, 11: \cite{Kraus2014}, 12:       \cite{Hoffleit1991}, 13: \cite{Kordopatis2013}, 13: \cite{Scholz2005}, 14: \cite{Lepine2013}, 15: \cite{Delorme2012}, 16: \cite{Gontcharov2006}, 17: \cite{Torres2006}}

\twocolumn

\begin{figure}[h]
\begin{center}
\includegraphics[width=0.40\textwidth]{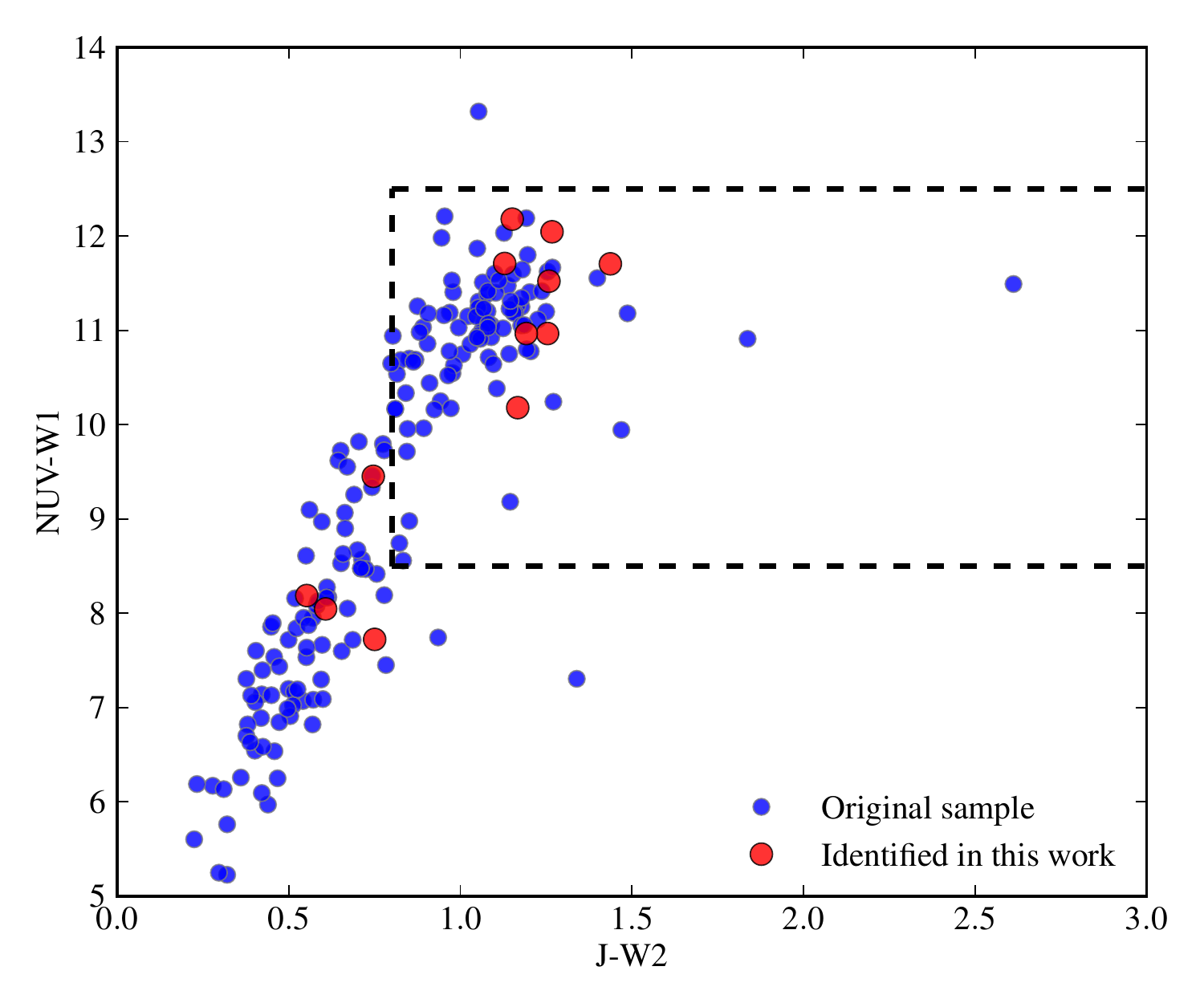}
\vspace{-0.2cm}
\caption{Colour-colour diagram for $J$-$W2$, NUV-$W2$ for new (red) and known (blue) members of the moving groups. The boxed area represents the criteria described in \cite{Rodriguez2013} used to classify potential young M dwarfs.}
\label{fig:nuv_comp}
\end{center}
\end{figure}

\cite{Rodriguez2011} combined near-IR photometry with UV emission to identify young, late-type (M0-M5) stars in Upper Scorpius and TW Hydrae.   The dotted-line box in Figure~\ref{fig:nuv_comp} shows the criteria imposed \cite{Rodriguez2013} to identify new members using NUV and infrared photometry.  We  plotted identified candidates and known members that have all the necessary photometry (NUV, $J$, $W1,$ and $W2$).  The few identified targets that have photometry in all four bands are either compatible, i.e. candidate young M dwarfs within the boxed area, or consistent with previous members of earlier spectral type.  Additionally, three of the targets matching this criteria have X-ray detections.  These sources have the largest L$_\mathrm{x}$/L$_\mathrm{bol}$ values shown in Figure~\ref{fig:xray_comp} (smaller than -3) for targets identified in this work. Objects matching the criteria are listed in Table~\ref{tab:identified_comps}.  The success rate of this technique, as derived in \cite{Rodriguez2013}, was $\approx$75\%.

\subsection{X-ray sources}
\label{sec:x_ray}

The original SACY sample was constructed from optical counterparts to X-ray bright sources from the ROSAT all-sky survey \citep[BSC;][]{Voges1999}.  We additionally queried the faint X-ray catalogue of ROSAT \citep[FSC;][]{Voges2000}, the serendipitous source catalogue of XMM \citep[3XMM;][]{XMM2013} and the Chandra source catalogue \citep[CSC; ][]{Evans2010S}.

\subsubsection{X-ray emission}

3XMM is the most recent serendipitous source catalogue from pointed observed using XMM-Newton.  Using Figure~8 of \cite{Watson2009} (2XMM catalogue) the lower limit of X-ray flux (0.2-2.0\,keV) ranges between 10$^{-15}$\,--\,10$^{-14}$\,erg\,cm$^{-2}$\,s$^{-1}$, depending on the sky coverage.  On average the XMM catalogue provides fluxes one order of magnitude fainter than ROSAT \citep{Voges1999}.  We cross-matched our detections with 3XMM for sources within 6\arcsec , and an upper search radius encompassed $>$99\% of positional uncertainties; see Figure~11 of \cite{Watson2009}.  We additionally searched the Chandra source catalogue \citep{Evans2010S} for sources within 5\arcsec using the CSCview\footnote{\url{http://cda.cfa.harvard.edu/cscview/}} programme to access data in the 0.2-2.0\,keV energy regime.  Our search yielded five sources with X-ray data (3: XMM, 1: FSC, 3: BSC, 1: CSC).

In the case of ROSAT data we converted the count rate and hardness-ratio (HR1) to X-ray flux using the formulation of \cite{Fleming1995}; these values are for energies in the range 0.1-2.4\,keV.  For CSC and XMM data the fluxes are available directly  in the catalogues, and we used the energies in the range 0.2-2.0\,keV to calculate total fluxes for an approximately equal comparison to fluxes derived from ROSAT data.  We converted the fluxes to luminosities assuming the distance to the target was that of its associated known member. 

\begin{figure}
\begin{center}
\includegraphics[width=0.49\textwidth]{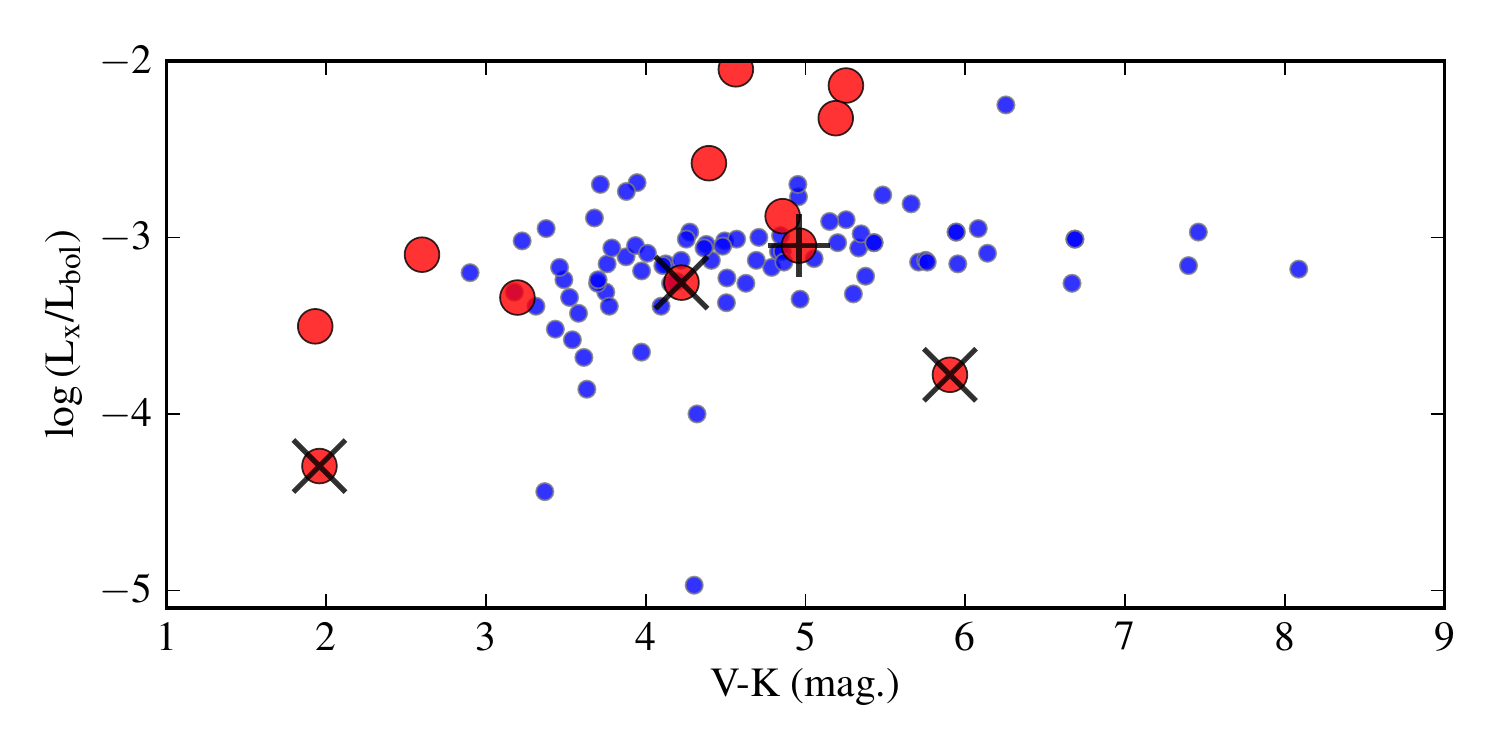}
\vspace{-0.2cm}
\caption{V-K colour versus log (L$\mathrm{x}$/L$_\mathrm{bol}$) for the original sample (blue) and for targets identified in this work (red).  The crosses represent targets identified as close ($<$0.3\arcsec) binaries; the luminosities are not corrected for unresolved components.  The plus symbols represent sources with a variability flag in the 3XMM catalogue. }\label{fig:xray_comp}
\end{center}
\end{figure}

We calculated bolometric luminosities of our targets using the VOSA tool \citep{Bayo2008}, which collates multi-band photometry from many reliable catalogues and fits both evolutionary models and black-body curves.  The values were calculated based on best-fit atmospheric models \citep{Allard2012} using available photometry and are listed in Table~\ref{tab:online_appendix_tab}.  For the nine members identified in this work with available X-ray data, we calculated the X-ray to bolometric luminosity ratio (L$\mathrm{x}$/L$_\mathrm{bol}$).  Three additional sources had a previously calculated values from \cite{Riaz2006} and \cite{Rodriguez2013}, which we also used. The results are presented in Figure~\ref{fig:xray_comp}.  

Typical values for young ($<$100\,Myr) sources are 10$^{-5}$ -- 10$^{-3}$ with saturation at $\approx$10$^{-3}$ \citep{Riaz2006, Zuckerman2004}.   One source (GJ 3305 AB) was identified previously as an X-ray bright, wide component (66\arcsec) to HD 29391 by \cite{Feigelson2006}.  The ratio we calculate in this work is in agreement with that derived previously.

Ten of the 12 sources are clearly consistent as they have ratios of -3.5 or lower.  The target at (5.9, -4.3) is a proposed member of AB Doradus (2MASS J23485048-2807157); given its colour and age (estimates range from 100-145\,Myr for the moving group) this target is still not ruled out as a young source from analysis of its X-ray emission (see Figure~4 of \cite{Zuckerman2004}).  Additionally it has strong lithium absorption and consistent radial velocity with its associated primary.  The remaining source (2MASS J05195327+0617258, at 2.0, -3.8) has a very low estimated mass (0.05\,M$_\odot$) and is a proposed member of the $\beta$-Pic moving group.  We did not find any additional information on this object, however its calculated ratio is still consistent with being a young member.

\subsubsection{X-ray non-detections}

Using the all sky ROSAT surveys \citep[RASS;][]{Voges1999, Voges2000} we also investigated non-detections based on expected L$\mathrm{x}$/L$_\mathrm{bol}$ values for young sources.  As shown in Figure~4 of \cite{Zuckerman2004}, young sources with a B-$V$ colour $>$0.7\,mag have L$\mathrm{x}$/L$_\mathrm{bol}$ values that are larger than -3.5. From our analysis (see Figure~\ref{fig:xray_comp}), we see the majority of sources also have values -3.5 or larger.  We used the bolometric luminosities calculated using VOSA \citep{Bayo2008} for the 50 sources in the colour range B-V$>$0.7\,mag. to calculate expected X-ray fluxes (\,$f_x=$L$_\mathrm{bol}\times10^{-3.5}$/$4\pi D^2$\,) based on this ratio.  The detection limit of RASS is $\approx$2$\times$10$^{-13}$\,erg\,s$^{-1}$ \citep{Schmitt1995}, 44/50 targets had predicted fluxes below this limit.  For the six sources with apparent non-detections we re-queried the RASS catalogues, extending the search radius beyond the positional error. Two of the six had nearby detections that were extremely extended, causing the positional offset of the source to lie within the radius of the detection, which could account for an apparent non-detection in this region.  We do not use the further four non-detections as a further youth constraint of the objects as one cannot be certain if these are reliable non-detections from the catalogue data.  For reference, the BSC catalogue has sky coverage of 92\% at a brightness limit of 0.1\,cts\,s$^{-1}$ \citep{Voges1999}.

\subsection{Gravity-sensitive features} 

\begin{figure}[h]
\begin{center}
\includegraphics[width=0.49\textwidth]{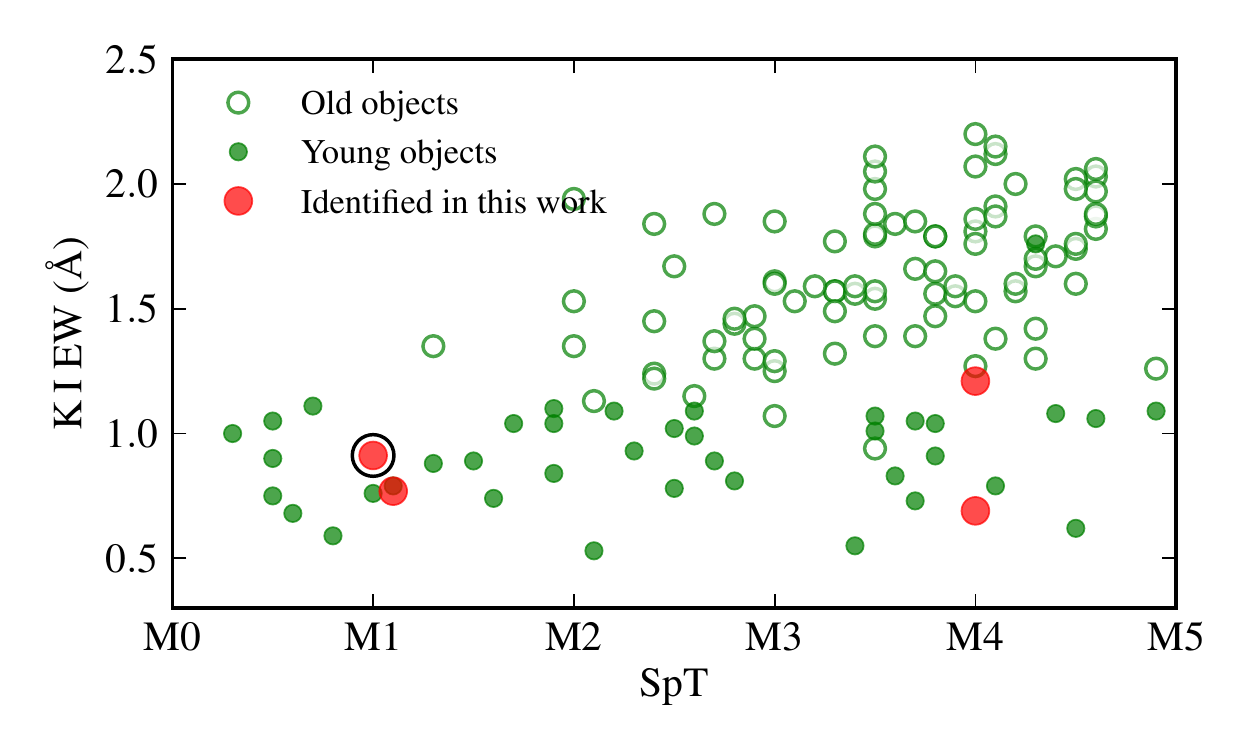}
\vspace{-0.2cm}
\caption{Spectral type versus EW of K\,I.  Green open and filled circles are for old and young objects from \cite{Shkolnik2011}, respectively.  Red filled circles are objects identified in this work.  The black open circle represents the RV-discrepant target 2MASS J02455260+0529240.}
\label{fig:grav_sens_features}
\end{center}
\end{figure}

We were able to calculate Na\,I and K\,I EWs for 11 objects from our spectroscopic observations. 
Figure~\ref{fig:grav_sens_features} shows the results for the EWs of K\,I for objects identified here and objects studied in \cite{Shkolnik2011} , which focussed on identifying nearby ($<$25\,pc) young M dwarfs.  We only show objects classified as M0 or later, the region where K\,I EWs distinguish between younger and older populations. All four objects have EWs consistent with youth, including the radial velocity-discrepant target 2MASS J02455260+0529240.

Given the S/N and wavelength coverage of the available spectroscopic observations, we could only estimate EWs of Na\,I (8194.8\,\AA) for spectral types earlier than $\approx$M2.  \cite{Slesnick2006} verified that the Na\,I strength saturates for spectral types earlier than M2.  Additionally the work of \cite{Schlieder2012b} verified that this feature is only useful for colours $V - K \geq 5$.  We therefore have not used this quantity in assessing the membership status of these objects.  The values are listed in Table\,\ref{tab:hermes_feros_obs} for reference.

\begin{figure}[h]
\begin{center}
\includegraphics[width=0.49\textwidth]{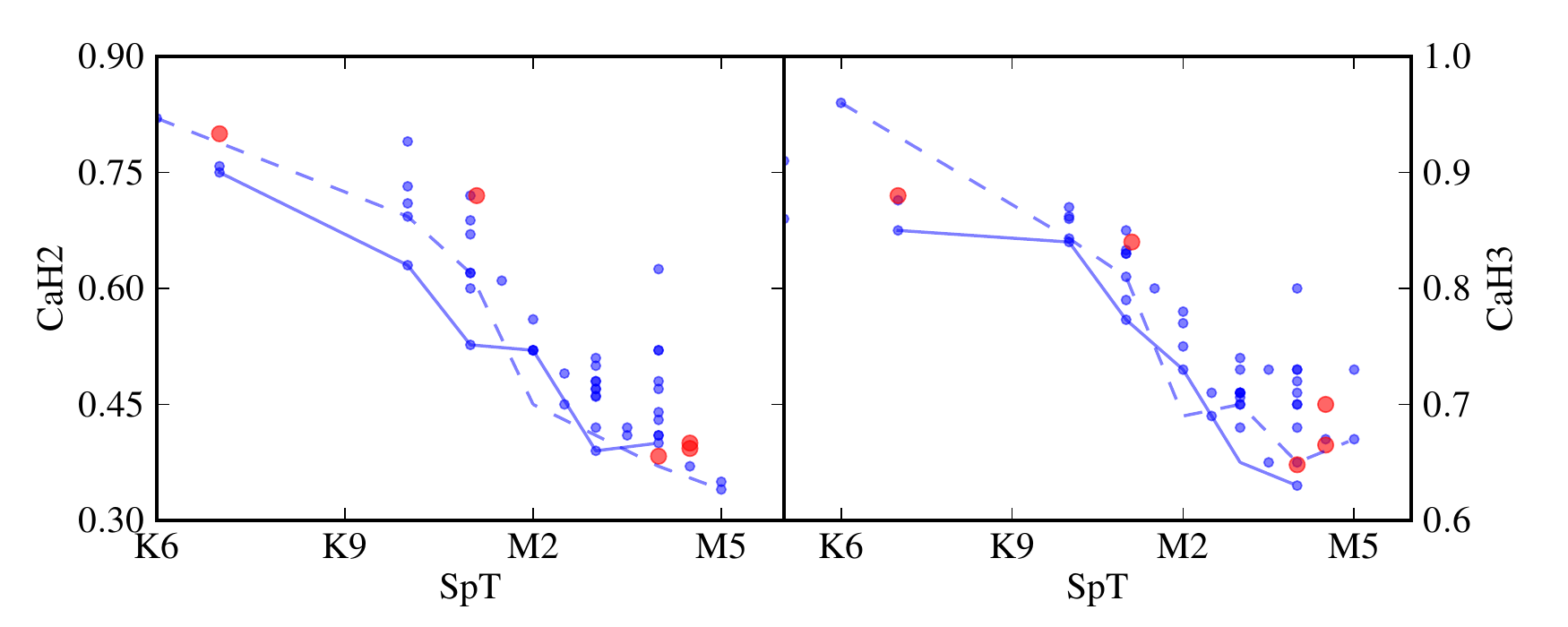}
\vspace{-0.2cm}
\caption{{\it Left panel:} spectral type versus CaH2 index for the original sample (blue) and members identified in this work (red).  {\it Right panel}: same as left, but for CaH3 index. The solid and dotted lines represent the lower envelopes of ABD and BPC, respectively.}
\label{fig:molec_bands}
\end{center}
\end{figure}

The visible spectra of M dwarfs are dominated by molecular bands (titanium oxide: TiO and calcium hydride: CaH).  \cite{Mould1976} showed that these bands are gravity-sensitive.  \cite{Reid1995} defined a series of band indices, including CaH2 and CaH3, to quantify the strength of these bands.  These indices measure the ratio of on-band to off-band flux.  We were unable to calculate these spectral indices using our own high-resolution spectra owing to low S/N.  However, we have collated values from other works the result of which are shown in Figure~\ref{fig:molec_bands}.  The dotted and solid lines represent the lower envelopes of indices for BPC and ABD, respectively.  There is a large amount of scatter for both indices ($\approx$0.1).  However, all the objects identified in this work have consistent CaH2 and CaH3 indices by comparison with known members.

Figure~\ref{fig:molec_bands} shows a comparison of CaH2 and CaH3 indices for the original sample and targets identified in this work.  The dotted and solid lines represent the lower envelopes of indices for BPC and ABD, respectively.  There is a large amount of scatter for both indices ($\approx$0.1).  All the objects identified in this work have consistent CaH2 and CaH3 indices by a comparison with known members.

\subsection{H$_\alpha$ emission} 

\begin{figure}[h]
\begin{center}
\includegraphics[width=0.49\textwidth]{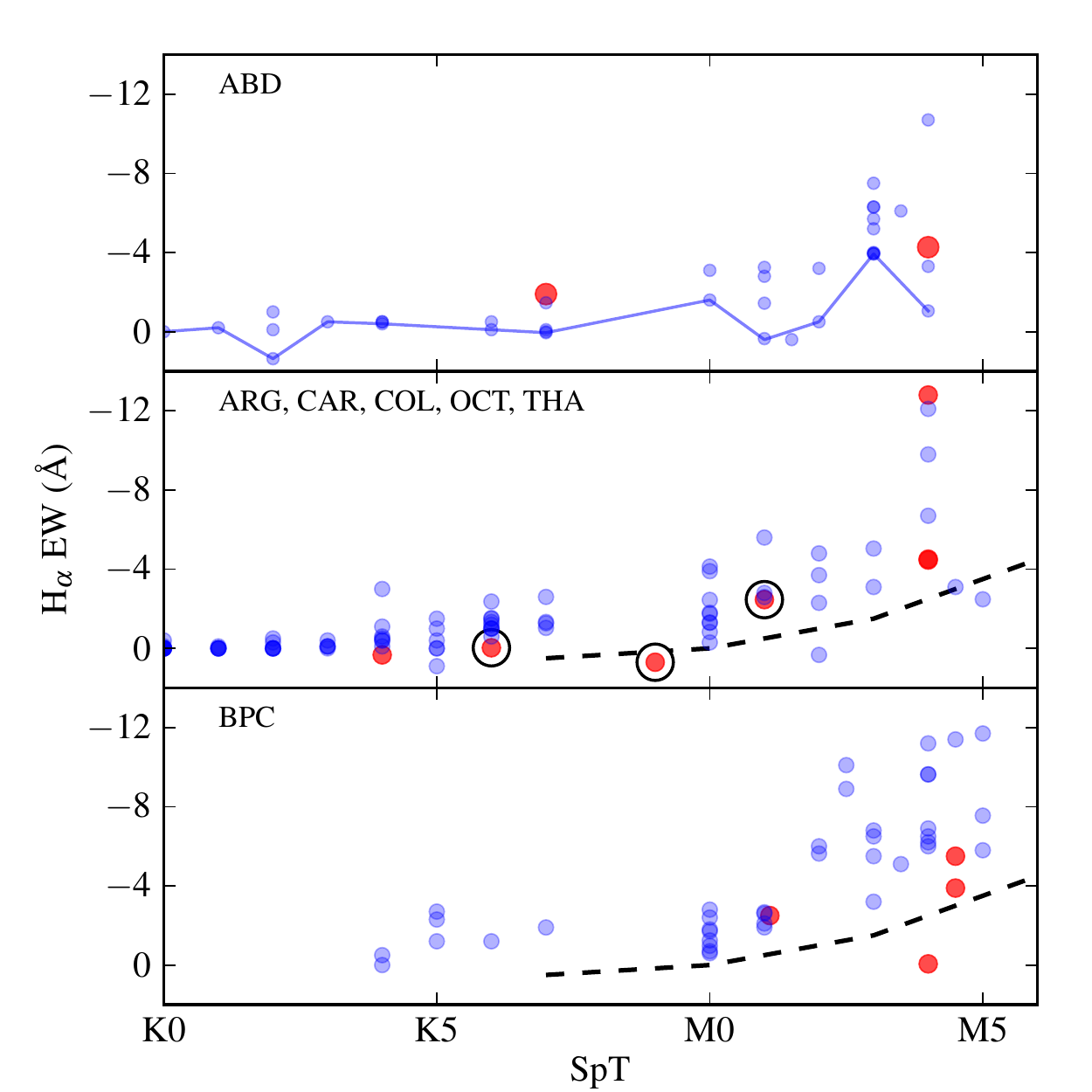}
\vspace{-0.2cm}
\caption{Spectral type versus H$_\alpha$ EW for the original sample (blue) and objects identified in this work (red). The black rings represent objects with discrepant radial velocities. The blue- and black-dotted lines are the lower envelopes of EWs for the original sample of ABD members and from \cite{Stauffer1997}, respectively.}
\label{fig:h_alpha}
\end{center}
\end{figure}

H$_\alpha$ emission can be used as an indication of youth, however, one has to keep in mind this feature can also be observed in older, main-sequence stars.  Furthermore, the EW of H$_\alpha$ is a function of the spectral type of the star at a given age, e.g. \cite{Stauffer1997}. In Figure~\ref{fig:h_alpha} we show the spectral type versus H$_\alpha$ EW for the original sample and targets identified in this work.  Additionally we show the lower envelope of EWs for the original sample of ABD and, similar to Figure~6 of \cite{Kraus2014}, show the lower envelope derived in \cite{Stauffer1997} (for the young open clusters IC 2391 and IC 2602, 25-35\,Myr).  We have grouped the objects by age (ARG, CAR, COL, OCT, THA all have age estimates $\approx$30-40\,Myr and are, hence, plotted together).   The envelopes act as an approximate lower boundary, as a function of age, to classify the object as having compatible H$_\alpha$ properties.  

There is  one object (2MASS J01354915-0753470) that has a discrepant EW, it is discussed in more detail in Appendix~\ref{sec:notes_individ_sources}.  

\subsection{Lithium abundance} 
\label{subsec:lithium}

If the mass of PMS star is $\approx$0.06\,M$_\odot$ or greater its core temperature eventually reaches $\approx$3\,MK and burns lithium \citep{Chabrier1996A}. The timescale to reach this temperature is mass dependent and once reached, the lithium is depleted rapidly as the mixing timescale is very short in convective stars.  

The lithium depletion boundary (LDB) method is based on observing at which luminosity, for a given population, this lithium burning occurs.  It is a very powerful technique to calculate the ages of young stars \citep{Soderblom2014}.  One of its main advantages is that it is less model-dependent than other techniques \citep{Bildsten1997, Jeffries2001}.  Lithium EWs can also provide accurate relative ages by comparing the abundance between different groups of stars \citep[see Figure~3 of][]{daSilva2009}. However, one must be cautious as even for young ages there is a large dispersion in the Li EWs that could be attributed to activity. \cite{Bayo2011} showed that measurements for the same object at different epochs and EWs between objects of the same spectral type can be significantly different.

\begin{figure}[h]
\begin{center}
\includegraphics[width=0.49\textwidth]{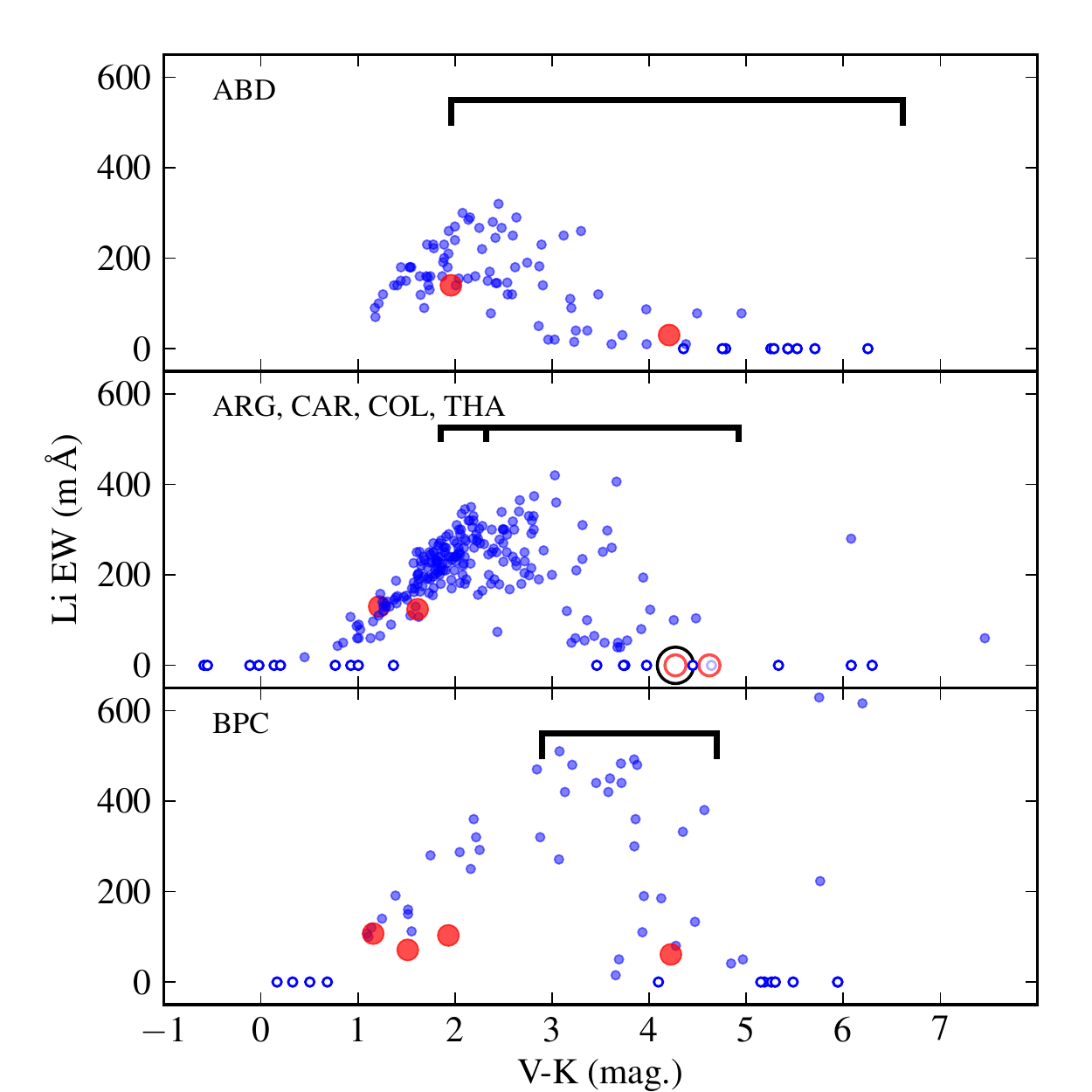}
\vspace{-0.2cm}
\caption{V-K magnitude versus Li EW for original sample (blue) and targets identified in this work (red).  Filled and open markers represent detections and non-detections, respectively.  Black open circles represent sources classified as non-members ("N") in Table~\ref{tab:identified_comps}.  The black lines encompass regions of lithium depletion (Li/Li$_0\leq$0.1) using the models of \cite{Baraffe2015}. In the middle panel there are two minima corresponding to ages of 30 and 40\,Myr.}
\label{fig:li_ew_comp}
\end{center}
\end{figure}

In Figure~\ref{fig:li_ew_comp} we plot the $V$-$K$ colour of the target versus the EW of lithium for 10 detections and non-detections in total. Additionally we show the region of lithium depletion (Li/Li$_0\leq$0.1) calculated using the evolutionary models of \cite{Baraffe2015}.

There are two ways to classify targets as having consistent lithium measurements; one from the empirical abundances from known members, the other from a comparison to the evolutionary models. All of our detections and non-detections are consistent with the models and empirical abundances.

{
\begin{table}
\tiny
\caption{Known members from our original sample that have been identified as wide companions to other known members.}
\begin{tabular}{p{2.5cm} p{0.7cm}  p{2.5cm} p{1.2cm}}
\hline\hline\\
  \multicolumn{1}{l}{ID} &
  \multicolumn{1}{l}{Dist.} &
  \multicolumn{1}{l}{(U, V, W)} &
  \multicolumn{1}{l}{Comps.} \\
  \multicolumn{1}{l}{} &
  \multicolumn{1}{l}{(pc)} &
  \multicolumn{1}{l}{(kms$^{-1}$)} &
  \multicolumn{1}{l}{} \\[1ex]  
\hline\\
 HD 104467 & 106.5 & -11.6, -18.9, -10.2 & SB\\
   GSC 09420-00948 & 110.5 & -11.6, -18.7, -10.3 & \\
\hline\\[-0.9ex]
 GSC 09239-01572 & 114.8 & -11.5, -19.9, -8.2 & 0.3\arcsec\\
   GSC 09239-01495 & 108.8 & -10.9, -21, -8.6 & \\
\hline\\[-0.9ex]
   HD 21423 & 107.5 & -13.0, -22.4, -5.2 & \\
   CD-36 1289 & 107.5 & -13.7, -21.1, -5.8 & \\
   HD 21434 & 107.8 & -12.2, -21.9, -5.7 & 0.2\arcsec\\
\hline\\[-0.9ex]
   GJ 140 & 16.9 & -24.3, -14, -6.5 & 3\arcsec\\
     GJ 140 C & 17.9 & -24.1, -13.9, -5.2\\
\hline\\[-0.9ex]
   HD 199058 & 76.0 & -6.2, -27.9, -13.3 & 0.5\arcsec\\
   TYC 1090-543-1 & 76.4 & -6.4, -27.2, -13.7 & \\
\hline\\[-0.9ex]
   TYC 112-1486-1 & 71.0 & -13, -15.4, -8.2 & 0.4\arcsec\\
      TYC 112-917-1 & 67.8 & -13.2, -15.5, -8.5 & \\
\hline\\[-0.9ex]
   Eta Tel & 47.7 & -9.1, -15.5, -8.3 & \\
     HD 181327 & 50.6 & -10.1, -16.4, -8.2 & \\
\hline\\[-0.9ex]
   HD 199143 & 45.7 & -7.9, -13.9, -10.9 & 1.1\arcsec \\
    AZ Cap & 47.9 & -10.1, -15, -9.5 & 2.2\arcsec \\
\hline\\[-0.9ex]
   HD 217343 & 32.0 & -2.9, -25.3, -14.2 & \\
     HD 217379 & 30.0 & -3.8, -27.5, -14.5  & 1.9\arcsec+SB\\
\hline\\[-0.9ex]
   HD 196982 & 9.5 & -9.8, -15.8, -9.6 & \\
     AU Mic & 9.8 & -10.8, -16.6, -9.9 & 2.9\arcsec\\
\hline\\[-0.9ex]
   UY Pic & 23.9 & -7.0, -28.1, -14.6 & \\
   CD-48 1893 & 24.2 & -7.0, -28.2, -14.6 & \\
\hline\\[-0.9ex]
   TYC 9300-891-1 & 147.6 & -12.0, -2.8, -9.5 & 1\arcsec  \\

   TYC 9300-529-1 & 147.0 & -11.5, -3.7, -9.6 &  \\
\hline\\[-0.9ex]
   BD-21 1074A & 20.2 & -11, -15.6, -9.4 & \\
   BD-21 1074B & 16.6 & -12.2, -15.3, -8.3 & 0.8\arcsec \\
\hline\\[-0.9ex]
 HD 13246 & 44.2 & -9.2, -20.3, -1.5 & SB\\
 CD-60 416 & 43.2 & -9.0, -20.5, -0.8 & \\
\hline\\
\end{tabular}
\label{tab:prev_identified_members}
\end{table}}

\subsection{Optical magnitudes}
\label{subsec:vmag_disc}

\begin{figure*}
\begin{center}
\includegraphics[width=0.95\textwidth]{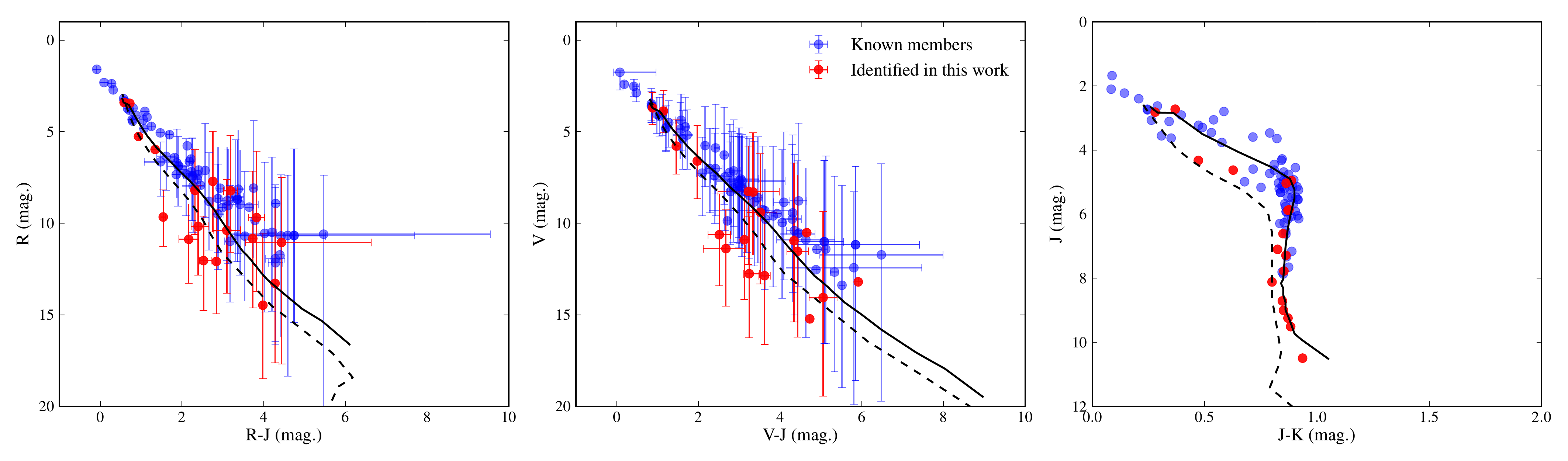}
\vspace{-0.2cm}
\caption{{\it Left and middle panel:} colour-magnitude diagrams ($R$-$J$, $R$) and ($V$-$J$, $V$) for known (blue) and newly proposed (red) members of the $\beta$-Pic association.  The uncertainties are derived using the upper and lower boundaries of the $R$ and $V$ magnitude values, respectively, compiled from our photometric search. In most cases those markers without uncertainties only have one magnitude value.  {\it Right panel:} colour-magnitude diagram ($J$-$K$, $J$) for known and newly proposed members.  The solid and dotted 
line is the isochrone of \cite{Baraffe2015} using the age of the association and the zero-age main sequence (ZAMS), respectively.}
\label{fig:hr_uncertainties}
\end{center}
\end{figure*}

From our compilation of optical photometric values, it is clear that catalogue-catalogue variations in both $V$ and $R$ magnitude can be very significant ($>2$\,mag.).  If these  magnitudes were used as a further mode of classification for our objects, we could be omitting viable candidates because of an apparent photometric mismatch, which is not necessarily associated with the sources membership status. These apparent variations in magnitudes could be induced by physical phenomena such as flaring \citep{Montes2005} at the time of observation from blending of multiple sources or purely poor measurement accuracy. Additionally, optical magnitudes are not available for all of our targets (for example, 18 sources have no $V$ mag., 35 sources have only one source of $V$ mag.). Figure~\ref{fig:hr_uncertainties} demonstrates how the variation in apparent optical magnitudes translates into large uncertainties in colour and absolute magnitude for existing and new members of the $\beta$-Pic moving group. Data points with no uncertainties only have one optical magnitude value. It is interesting to note that many of the identified members lie closer to the zero-age main sequence (ZAMS) than to their respective isochrone.  This could be an indication that some of the identified targets in this work are from the older field population.  Further spectroscopic observations will allow us to conclude whether these targets are young.  
Colour-magnitude diagrams of the remaining eight associations are shown in Appendix~\ref{sec:app_hr_diagrams}. 

As mentioned in Section~\ref{sec:catalina_phot} we checked all sources with Catalina photometry for any signs of optical flaring that could be associated with variable strong activity \citep{Montes2005}.  We found 4 / 36 identified targets in this work, with Catalina photometry, showed significant flaring.  We classify significant flaring as two or more observations with values three times the standard deviation of all data.  None of these four objects have optical spectra to connect the photometric flare with strong, variable activity. For that reason, this quantity is not used to assess the membership of the target but, it is noted in Table~\ref{tab:identified_comps}.

\subsection{Revised mass function of the young associations}

As mentioned previously, the original works aiming to derive censuses of the young associations suffered from large incompleteness for masses $<$1\,M$_\odot$.  However, recent work has significantly improved members in this mass range. 

\begin{figure}[h]
\begin{center}
\includegraphics[width=0.45\textwidth]{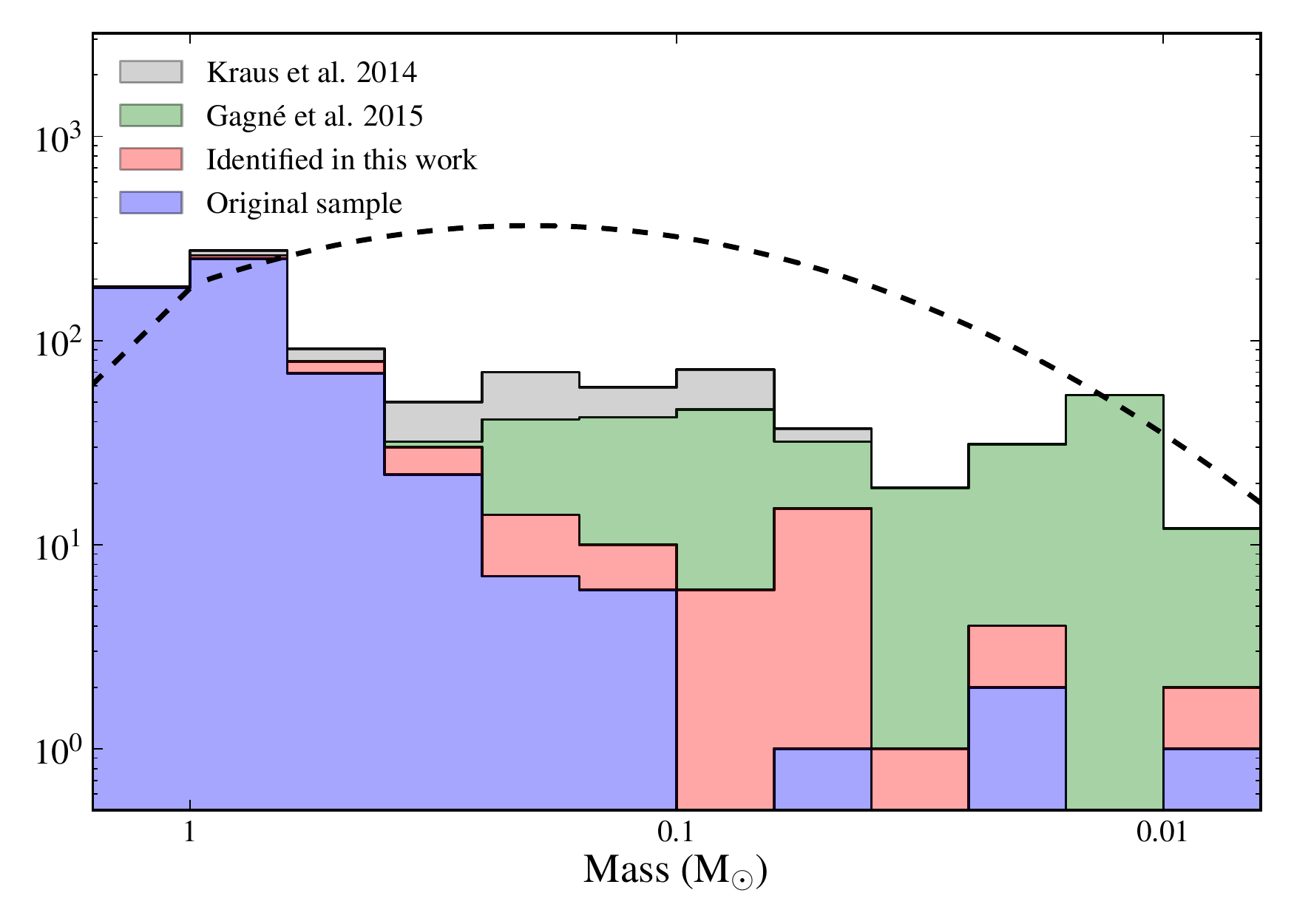}
\vspace{-0.2cm}
\caption{Mass function of the nine young moving groups for objects identified in this analysis (red), \cite{Gagne2015} (green), \cite{Kraus2014} (grey) and the original sample (blue).  A power law is fitted for masses $>1$\,M$_\odot$ \citep[$\alpha$=-2.35,][]{Salpeter1955} and a log-normal ($\mu$=0.2\,M$_\odot$, $\sigma$=0.6\,log(M$_\odot$)) for masses smaller than 1\,M$_\odot$ \citep{Chabrier2003}.  }
\label{fig:imf_estimate}
\end{center}
\end{figure}

Figure~\ref{fig:imf_estimate} shows the IMF, when considering all nine young associations as one overall population.  This figure includes targets from the original sample, targets identified in this work (a pseudo-random sample of all 84 targets multiplied by the success rate), high-probability targets from \cite{Gagne2015} and {\it consistent} targets (based on RV, H$_\alpha$ and/or Li) from \cite{Kraus2014}. Although the work of \cite{Kraus2014} was only focussed on the Tucana-Horologium association, this work significantly increased the number of potential members from 62 to 191; therefore even in the context of all nine associations still makes a large difference to the IMF (see the log-scale of Figure~\ref{fig:imf_estimate}).  

The most considerable improvement from analysis presented here  is for stars with masses $\lesssim$0.5\,M$_\odot$.  We are sensitive to low-mass objects as our initial selection criteria is only based on 2MASS photometry and proper motions.  Many other works, such as \cite{Rodriguez2013}, use more extensive selection criteria, but at the cost of severely limiting the number of sources with photometry in multiple bands.  This is highlighted by Figure~\ref{fig:nuv_comp}, which shows only 13 of 84 sources (15\%) had photometry in both GALEX and WISE.

\section{Conclusions}
\label{sec:conclusions}

We have identified 84 targets (43:\,0.2-1.3\,M$_\odot$, 16:\,0.08-0.2\,M$_\odot$, 25:\,$<$0.08\,M$_\odot$)  using our sample of 542 high-probability members.  Thirty-three out of 84 sources have spectroscopic parameters either derived here or taken from previous works listed in Table~\ref{tab:prev_surveys}.  Of these 33 sources, four sources have inconsistent parameters and four have questionable parameters.  We can therefore derive a success rate of $\approx$ 76--88\% using this technique.  Additionally only ten out of 33 have been identified in previous works concerned with young associations.  Thanks to the collation of numerous parameters, we were able to find and use existing additional information from mid- to high-resolution spectra and X-ray observations to successfully complement our photometric--proper motion search for new members.  The targets with existing additional information, on which the success rate is derived, are generally brighter in magnitude (average in $V$: 12.2\,mag. compared to 15.1\,mag) and earlier spectral type (average K7 compared to M0) than those currently without further photometric and spectroscopic information. Therefore the derived success rate is only an approximation at this time and should be treated in that way.  In the future we aim to take mid- to high-resolution spectra of all the targets identified in this work that currently have no spectroscopic information.  With this approach, we can better assess the youth and kinematics of the objects.  \\

As expected by our success rate so far, if many of these new members are confirmed they could have big implications for the formation and dynamical evolution of stars in loose associations.  Our initial results are broadly consistent with the mechanism proposed by \cite{Reipurth2012} whereby preferentially lower mass components in three-body systems are ejected from the stellar cores early on their lifetime.  These components collectively form {\it unfolding} triple systems that are potentially in the process of disintegrating into unbound systems.  The majority of these wide companions would therefore be associated with tight inner binaries (tightening of the inner binary from angular momentum exchange as the third body is ejected).  An evaluation of this mechanism using the young associations will be explored in further work (Elliott et al. in prep.).  

Regardless of whether these targets form wide multiple systems or not, if their youth is confirmed, they would make up an ideal sample for the characterisation of low-mass objects and the search for warm planets around nearby stars.    New instruments such as SPHERE/VLT \citep{Beuzit2008} and GPI/GEMINI \citep{Macintosh2014} have been designed for the purpose of detecting and characterising planets using high-contrast imaging techniques.  For a strehl ratio of $\approx$70\% or better in the $H$ band, the limiting magnitudes for SPHERE and GPI are $R<$13\,mag and I$<$10\,mag, respectively.  From our sample of identified candidates in this work, 24 have $R$ magnitudes brighter than the limit of SPHERE. Fifteen of these have been classified as members (denoted by "Y" in Table~\ref{tab:identified_comps}) based on further photometric and spectroscopic quantities, and nine of these 15 are M-dwarfs.  We have therefore, based on this initial work alone, identified crucial new targets to search for warm planets around nearby young stars.

\begin{acknowledgements}
PE would like to thank A. Tokovinin for a visit in August 2015 that significantly helped to improve the data management of this project, in part making the analysis presented here possible. PE is also grateful to I. Baraffe for a finer grid of models upon request. A. Bayo acknowledges financial support from the Proyecto Fondecyt de Iniciación 11140572. D. Montes acknowledges financial support from the Spanish Ministerio de Economía y Competitividad 
under grants AYA2011-30147-C03-03 and AYA2014-54348-C3-3-R.  This research has made use of the Washington Double Star catalogue maintained at the U.S. Naval Observatory, the Topcat tool \citep{Taylor2005}, the SIMBAD database, and the VizieR catalogue access tool, CDS, Strasbourg, France (the original description of the VizieR service was published in A\&AS143, 23)
We have used Catalina (CSS) data in our analysis. The CSS survey is funded by the National Aeronautics and Space Administration under Grant No. NNG05GF22G issued through the Science Mission Directorate Near-Earth Objects Observations Program.  The CRTS survey is supported by the U.S.~National Science Foundation under grants AST-0909182 and AST-1313422.  This publication makes use of VOSA, developed under the Spanish Virtual Observatory project supported from the Spanish MICINN through grant AyA2011-24052.
\end{acknowledgements}

\bibliography{/Users/pelliott/Documents/PhD/LaTex_Papers/LaTex/biblio1}


\begin{appendix}

\section{Notes on compatibility of individual sources}
\label{sec:notes_individ_sources}

Identified targets with discrepant or dubious properties (denoted by either "N" or "Y?" in Table~\ref{tab:identified_comps}) are discussed below. We combine all available indicators of youth in the discussion to assess their potential membership. Uncertainties of radial velocity values quoted are 1.0\,kms$^{-1}$ in the case of measurements performed by the SACY team.  This is not a measurement uncertainty but rather it is the average variation seen in single stars from using the CCF technique on optical spectra \citep[see][for details]{Elliott2014}.\\

\noindent{\it 2MASS J02105345-4603513}: The radial velocity of this target with its associated known member are incompatible 25.7$\pm$1 and 8.5$\pm$1\,kms$^{-1}$.  However, the evidence that these two sources form a young wide binary is strong.  It is identified as a multiple system in the WDS catalogue (ID: 02109-4604, observations 1977-2000), \cite{McCarthy2012}, and \cite{Rodriguez2013}.  However, because of the angular separation (21.5\arcsec) there is no observable orbital motion on this timescale.  The proper motion values match within 1\,sigma (total proper motions: 54.4$\pm$3.1 and 54.2$\pm$2.5\,mas/yr) and additionally the photometry is compatible in $V$, $J$, $H$ and $K$.  2MASS J02105345-4603513 also shows strong X-ray emission (L$_\mathrm{x}$/L$_\mathrm{bol}$=-2.88) and UV excess.  It is very likely that 2MASS J02105345-4603513 is a spectroscopic binary and the radial velocity value calculated by \cite{Rodriguez2013} is not the system velocity.  For those reasons we conclude 2MASS J02105345-4603513 is a most likely a young source.\\


\noindent{\it 2MASS J12573935+3513194}: The radial velocities of this target with its associated known member borderline agree (2-3\,$\sigma$) within measurement uncertainties (-9.5$\pm$1.0 and -5.7$\pm$0.5\,kms$^{-1}$), however both components have been identified as spectroscopic binaries.  The separation between the two components is relatively small (16.2\arcsec) and the photometry matches well for $V$, $J$, $H,$ and $K$. 2MASS J12573935+3513194 shows strong H$_\alpha$ emission (-4.27), indicating it is young. Additionally the proper motion of the components are both extremely high and agree within uncertainties (total proper motion: 303.9$\pm$2.8 and 317.8$\pm$7.6\,mas/yr).   \\


\noindent{\it 2MASS J01334282-0701311}: Initially identified as part of a triple system.  The components have radial velocities -15$\pm$0.1 (HR 448), 11.4$\pm$0.2 (2MASS J01334282-0701311) and 12.2$\pm$1\,kms$^{-1}$ (G 271-110), respectively.  The lowest mass component (G 271-110) in this system is the original known member from \cite{Malo2014}.  The highest mass component (HR 448) is a well-documented target with extensive radial velocity coverage and therefore this discrepancy is unlikely to arise from variation induced by spectroscopic companion.  
\\

\noindent{\it 2MASS J21212446-6654573}: There are two epochs of data for this target, however one radial velocity has huge uncertainties (-24.1$\pm$1 and 6.4$\pm$14.8\,kms$^{-1}$).  \cite{Torres2006} also noted this component as a potential spectroscopic binary.  The proper motions do not agree within 3$\sigma$ however, the magnitudes of the values are very large, 142.4$\pm$1.9 and 128.3$\pm$1.9\,mas/yr, respectively. The field of view is not crowded and positions of the targets are fully outside the Galactic plane (Galactic longitude and latitude: 326.596, -39.1878) making a chance alignment improbable.  The system is also identified as a multiple system (with no additional notes) in the WDS catalogue, with seven observations between 1835-2000. The photometry of both components matches very well in $V$, $J$, $H$ and $K$. \\

\noindent{\it 2MASS J04515223-4640497}: This target has a hugely discrepant radial velocity with its associated known member (51.2$\pm$1.0\ and 23$\pm$1.0\,kms$^{-1}$). It has small emission in H$_\alpha$ (EW: -0.32\,\AA).  As a result of its discrepant radial velocity, this target is not classified as a potential member.\\

\noindent{\it 2MASS J08371456-5517343}: This target has a hugely discrepant radial velocity with its associated known member (81.3$\pm$1.0\ and 20.8$\pm$1.0\,kms$^{-1}$) and low proper motion magnitude (12.5$\pm$2.6\,mas/yr). It has absorption in H$_\alpha$ therefore there is no evidence it is a young member. \\

\noindent{\it 2MASS J01354915-0753470}: This target has consistent radial velocity (6.3$\pm$0.5 and 6.5$\pm$1.0\,kms$^{-1}$) and very large and consistent proper motion (104.7$\pm$2.8 and 98.1$\pm$7.2) with its associated known member, both agree within 1\,sigma. The calculated K\,I EW is consistent with youth, however it does not show H$_\alpha$ emission. 
Given its V-K colour this makes the youth of the object questionable. \\

\noindent{\it 2MASS J02455260+0529240}: The calculated radial velocity is hugely discrepant with its associated known member (88.5 and 4.3\,kms$^{-1}$) and there is no sign of a spectroscopic companion to account for such a discrepancy.  The total proper motions are large and agree within 1 sigma (83.0$\pm$4.2 and 87.6$\pm$0.9\,mas/yr).  However, due to RV value, it is not considered as a potential member.

\section{Previously identified extremely wide companions}
\label{sec:pre_identified_wide_comps}

Below we present previous detections of potential wide binary systems in the young associations.  We discuss whether these systems were recovered in the analysis presented here and the reasons for why or why not. \\

\begin{figure}[h]
\begin{center}
\includegraphics[width=0.45\textwidth]{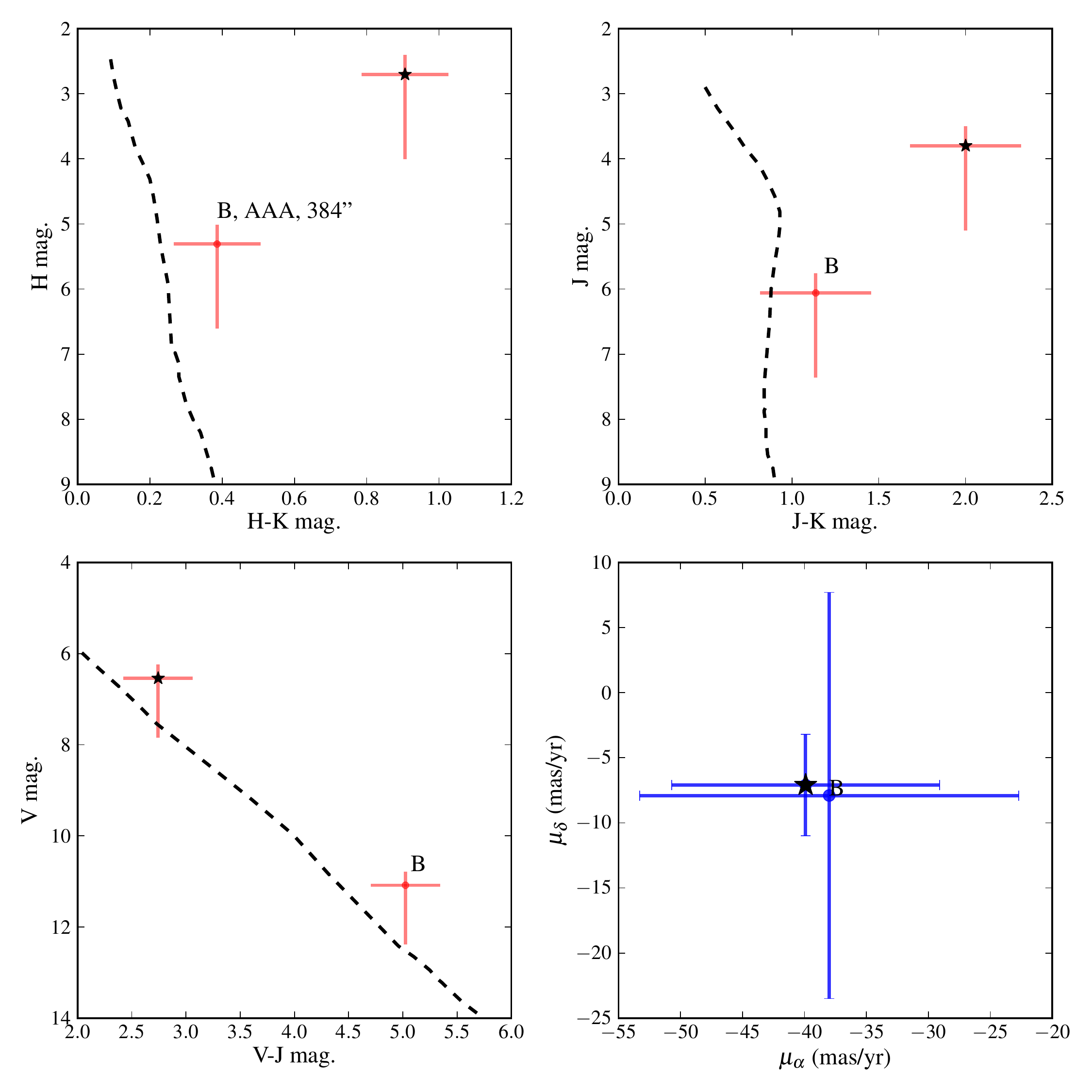}
\vspace{-0.2cm}
\caption{Photometric and kinematic summary of 2MASS J11550485-7919108.}\label{fig:tcha_hr_diagram}
\end{center}
\end{figure}

\textit{T Cha -- 2MASS J11550485-7919108}: This system was first identified in \cite{Kastner2012}.  In our analysis we did not recover the system as the H-K colour is incompatible ($<$0.03\,mag. beyond criterion, see Figure~\ref{fig:tcha_hr_diagram}).  
As discussed previously at the youngest ages ($\approx$10\,Myr) our technique is weaker as infrared excesses are more likely and therefore the companion is less likely to be classified as possibly physical.  However, this effect is unlikely to induce any strong bias in our statistics.  \cite{Kastner2012} show 2MASS J11550485-7919108 has a modest infrared excess ($K$-$W3\sim$0.8) which is why it was not recovered here.  The system is classified as a potential wide binary system because of the detailed analysis performed by \cite{Kastner2012}, including H$_\alpha$, X-ray and lithium EW analysis, and the recent work of \cite{Montet2015}.  \\

\begin{figure}[h]
\begin{center}
\includegraphics[width=0.45\textwidth]{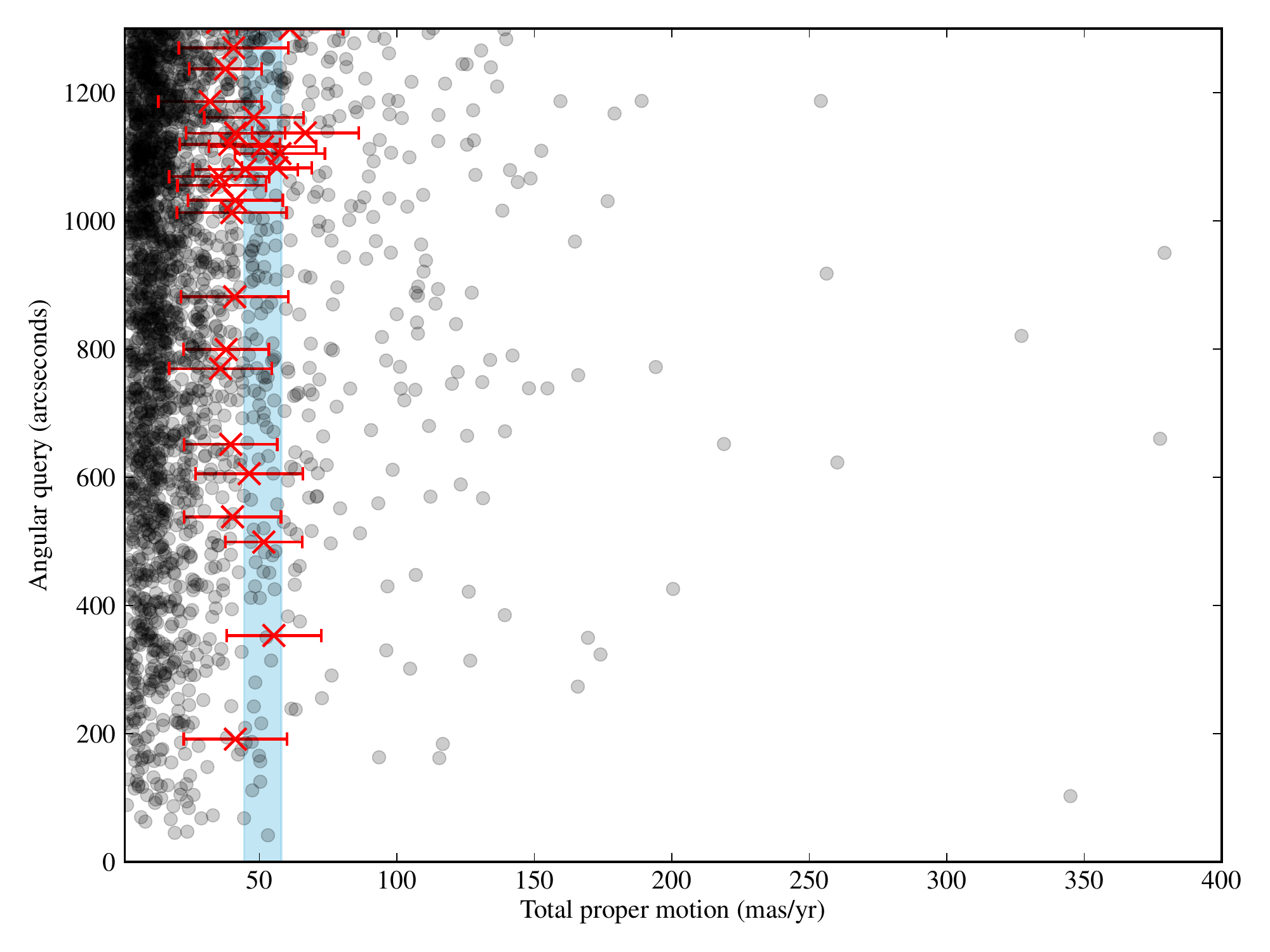}
\vspace{-0.2cm}
\caption{Considered field of view for V4046 Sgr.  The markers are the same as in Figure~\ref{fig:contamination}}
\label{fig:v4046_contam}
\end{center}
\end{figure}

\noindent\textit{V4046 Sgr -- GSC 07396-00759}:  First identified in \cite{Kastner2011}, this companion does have compatible photometry and kinematics using our method.  However, many other sources in the field of view also match the criteria (see Figure~\ref{fig:v4046_contam}) and therefore our upper angular limit is set much smaller (10\arcsec) than the angular separation of this candidate companion (2.82\arcmin). It is also noteworthy that the kinematic distances are not compatible between the two sources 76.9 and 95.1\,pc, respectively.  Furthermore \cite{Kastner2011} argue that GSC 07396-00759 would itself have to be a spectroscopic multiple system (of approximately equal mass) to account for the discrepancy in magnitudes.  However, spectroscopic analysis does not support this.  From 3 separate epochs of data there is no evidence for a companion.  The radial velocities are -5.7, -5.0 and -5.7\,kms$^{-1}$ ($\sigma$=0.3\,kms$^{-1}$) from \cite{Torres2006} and this work, which is well below the criterion for multiplicity ($\sigma>$3\,kms$^{-1}$). For these reasons the system is not included in our statistics.

\begin{figure}[h]
\begin{center}
\includegraphics[width=0.45\textwidth]{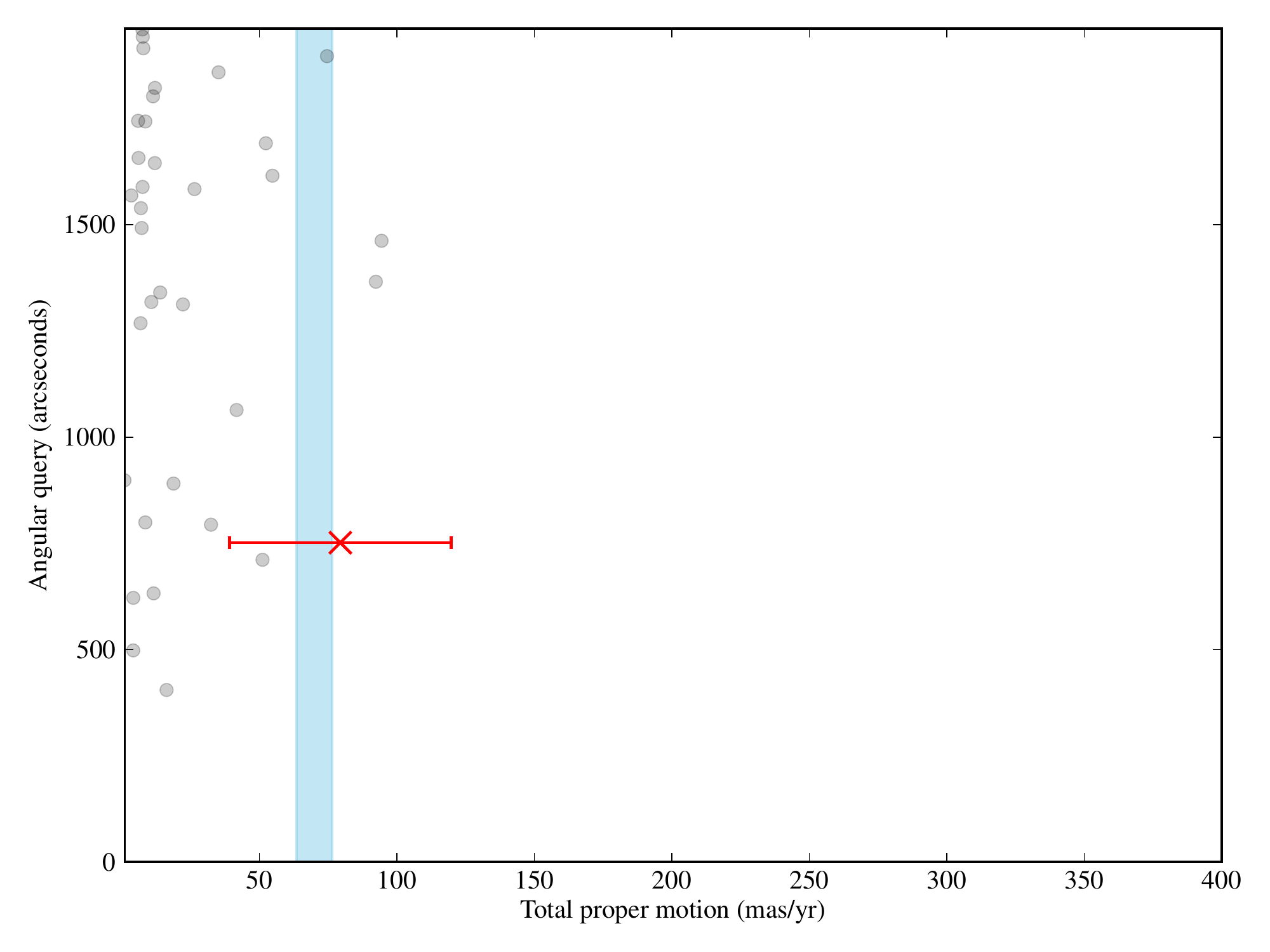}
\vspace{-0.2cm}
\caption{Considered field of view for TWA 1.  The markers are the same as in Figure~\ref{fig:contamination}. 2MASS J11020983-3430355 is the only source with compatible proper motion.}
\label{fig:twa1_contam}
\end{center}
\end{figure}

\noindent\textit{TWA 1 -- 2MASS J11020983-3430355}: This system was first identified in \cite{Scholz2005}.  The system is also recovered here.  Figure~\ref{fig:twa1_contam} shows that 2MASS J11020983-3430355 is the only source with both compatible photometry and kinematics in the field of view. Furthermore the work of \cite{Scholz2005} used spectroscopic analysis to show the source is consistent with being a member of the TW Hydrae moving group. The system is therefore included in our statistics.\\

\noindent\textit{TWA 30 -- 2MASS J11321822-3018316}:  This system was first identified in \cite{Looper2010}.  The system was also recovered here, however, initially it was discarded due to the extremely large uncertainties on the proper motion (14.3\,mas/yr).  The work of \cite{Looper2010} produced much more accurate proper motion values (uncertainties 9\,mas/yr) and a radial velocity value (12$\pm$3\,kms$^{-1}$), showing its Galactic velocity is compatible with that of TWA 30 -- (-10.2, -18.3, -4.9) compared with (-11.7, -19.7, -4.1) for TWA 30.   Furthermore their spectroscopic analysis showed the source has features consistent with being a young object, member of the TW Hydrae moving group.  For these reasons it is included in our statistics. \\

\noindent\textit{51 Eri -- 2MASS J04373746-0229282}: This system was first identified in \cite{Feigelson2006} and is also recovered in our analysis.  The work of \cite{Feigelson2006} concludes that these two objects are very likely a physical pair.  In recent work \cite{Montet2015} calculated the system velocity of 2MASS J04373746-0229282, itself a close binary, as 20.8$\pm$0.2\,kms$^{-1}$.  The radial velocity of 51 Eri is 21.0$\pm$1.2\,kms$^{-1}$, therefore the two components of the wide binary system are consistent.

\newpage
\onecolumn

\section{Colour-magnitude diagrams for new candidates}
\label{sec:app_hr_diagrams}

Below are colour-magnitude diagrams (in the same format as Figure~\ref{fig:hr_example}) for the remaining eight associations.

\begin{figure}[h]
\begin{center}
\includegraphics[width=0.95\textwidth]{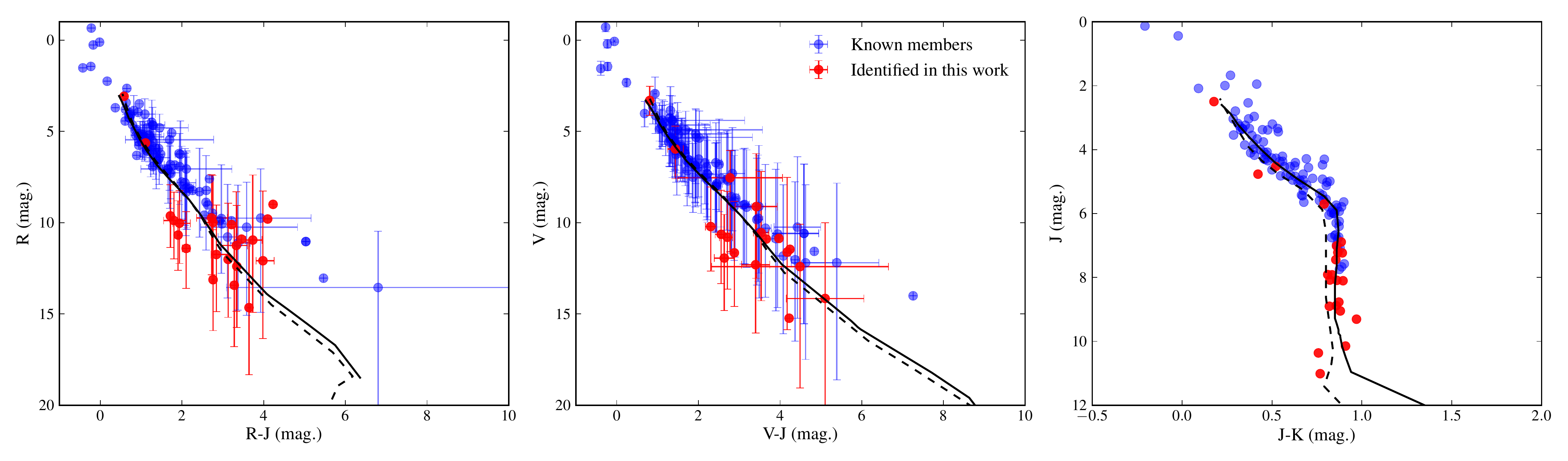}
\vspace{-0.2cm}
\caption{Colour-magnitude diagrams for the AB Doradus association}
\label{fig:hr_abd}
\end{center}
\end{figure}

\begin{figure}[h]
\begin{center}
\includegraphics[width=0.95\textwidth]{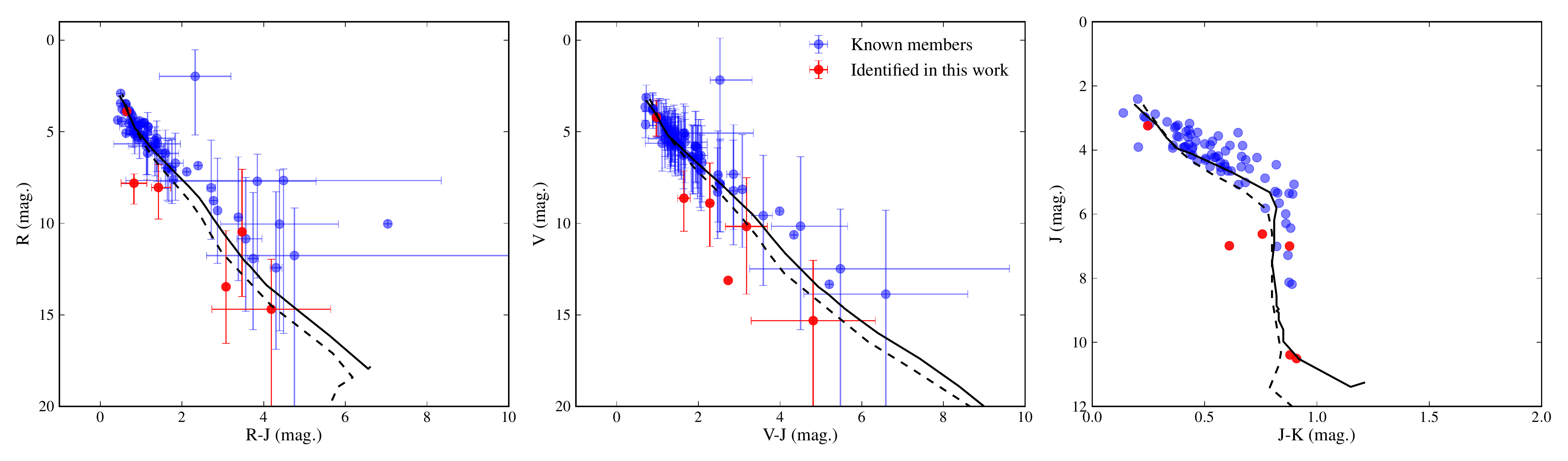}
\vspace{-0.2cm}
\caption{Colour-magnitude diagrams for the Argus  association.}
\label{fig:hr_arg}
\end{center}
\end{figure}

\begin{figure}[h]
\begin{center}
\includegraphics[width=0.95\textwidth]{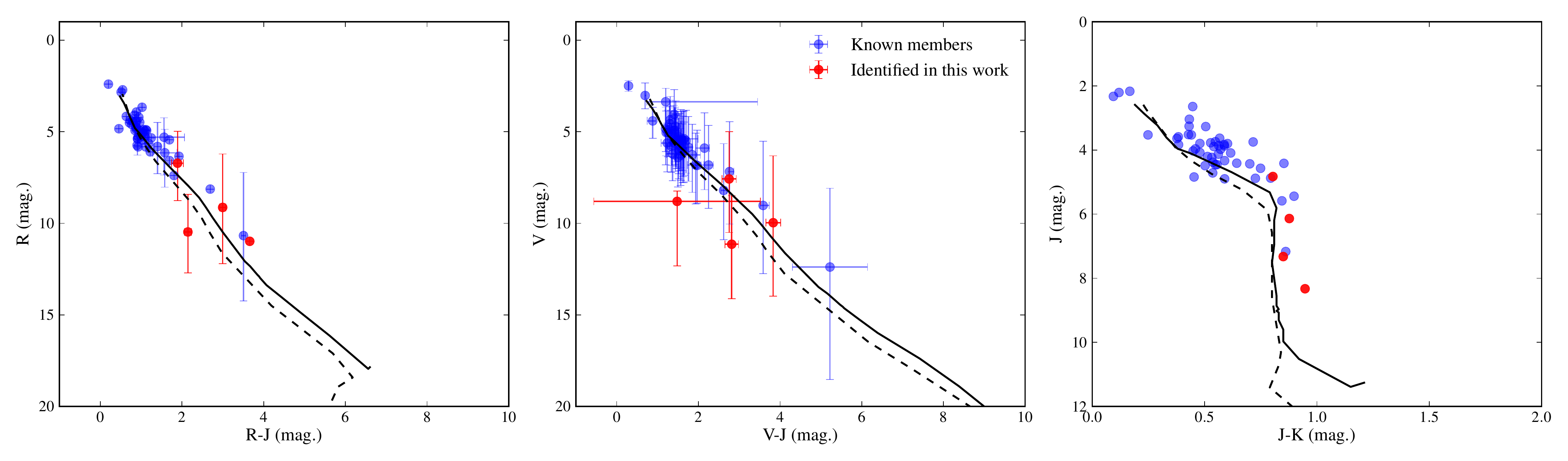}
\vspace{-0.2cm}
\caption{Colour-magnitude diagrams for the Carina association.}
\label{fig:hr_car}
\end{center}
\end{figure}

\begin{figure}
\begin{center}
\includegraphics[width=0.95\textwidth]{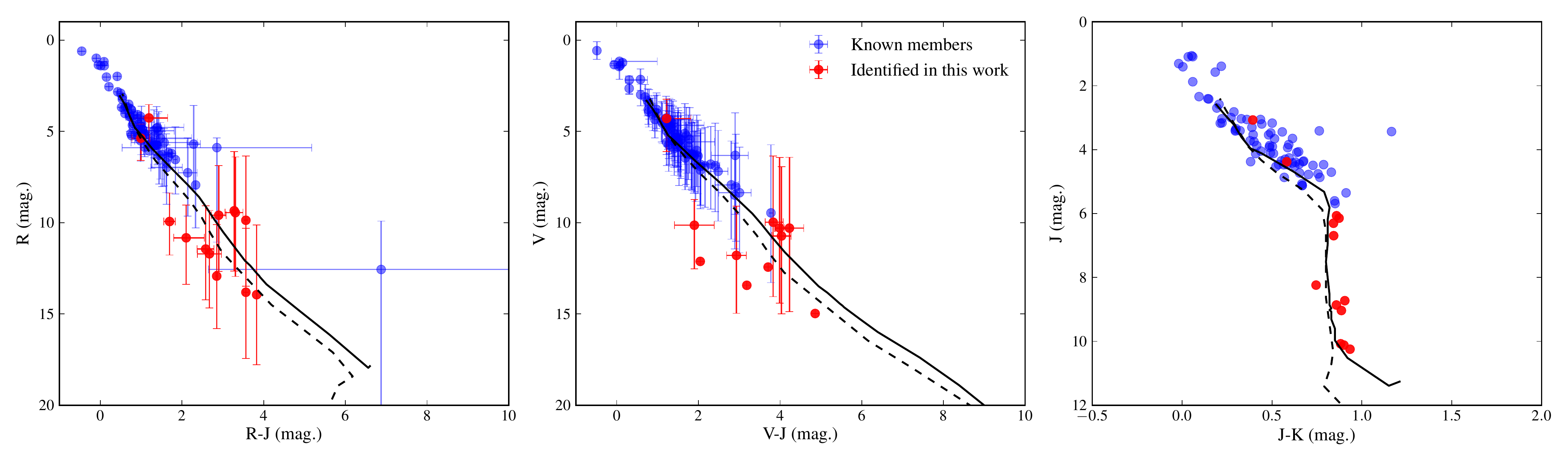}
\vspace{-0.2cm}
\caption{Colour-magnitude diagrams for the Columba association.}
\label{fig:hr_col}
\end{center}
\end{figure}

\begin{figure}
\begin{center}
\includegraphics[width=0.95\textwidth]{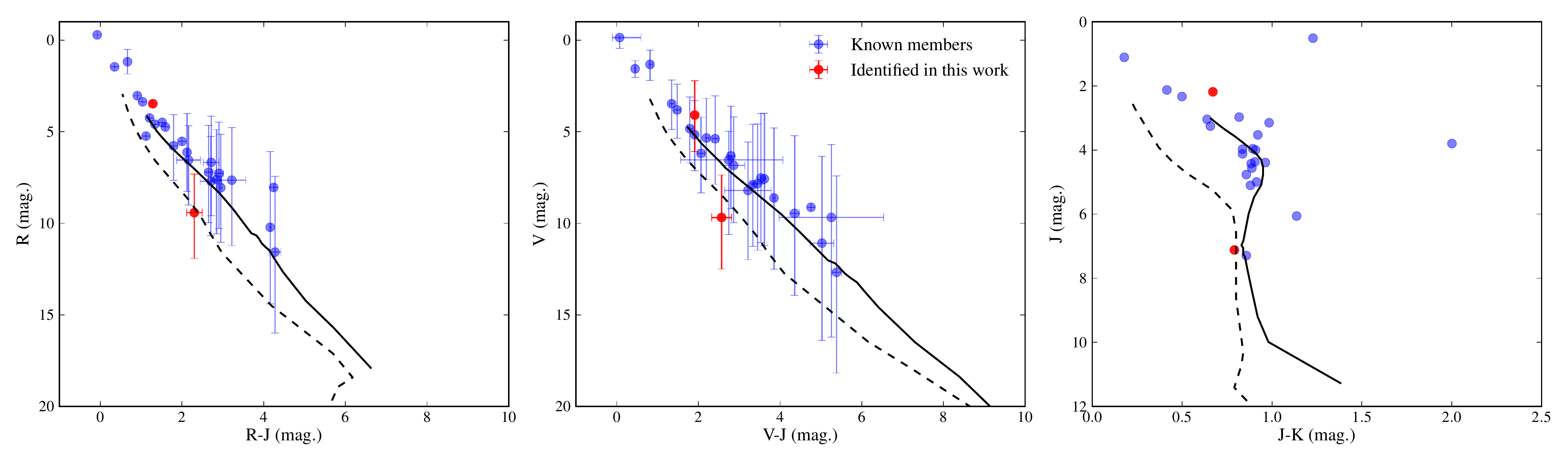}
\vspace{-0.2cm}
\caption{Colour-magnitude diagrams for the $\epsilon$-Cha association.}
\label{fig:hr_ech}
\end{center}
\end{figure}

\begin{figure}[h]
\begin{center}
\includegraphics[width=0.95\textwidth]{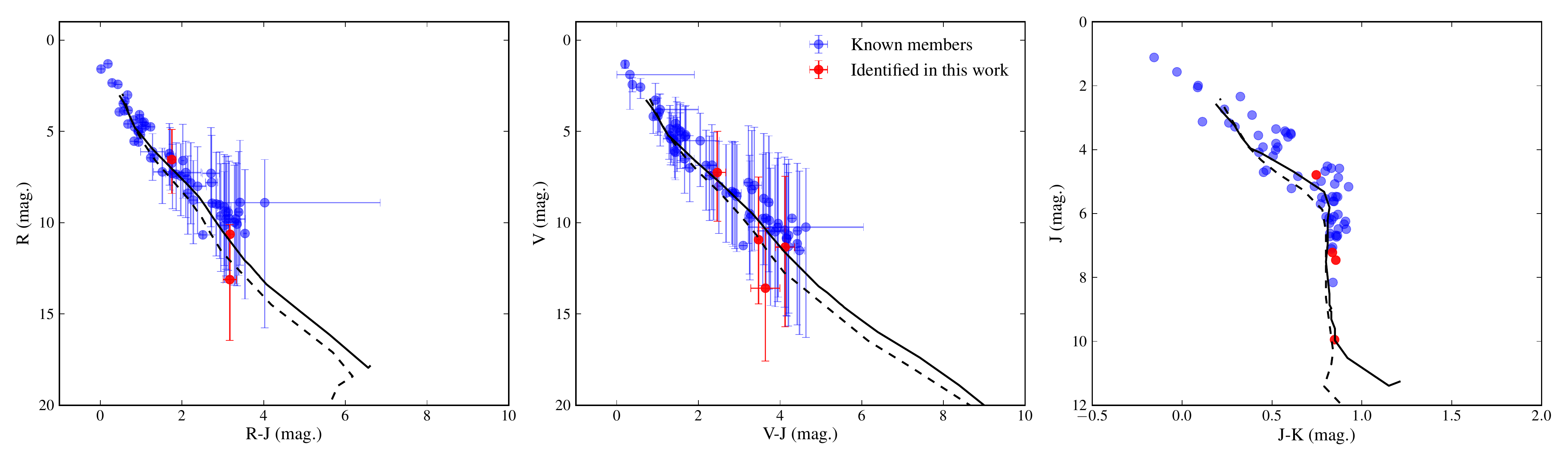}
\vspace{-0.2cm}
\caption{Colour-magnitude diagrams for the Octans  association.}
\label{fig:hr_oct}
\end{center}
\end{figure}

\begin{figure}
\begin{center}
\includegraphics[width=0.95\textwidth]{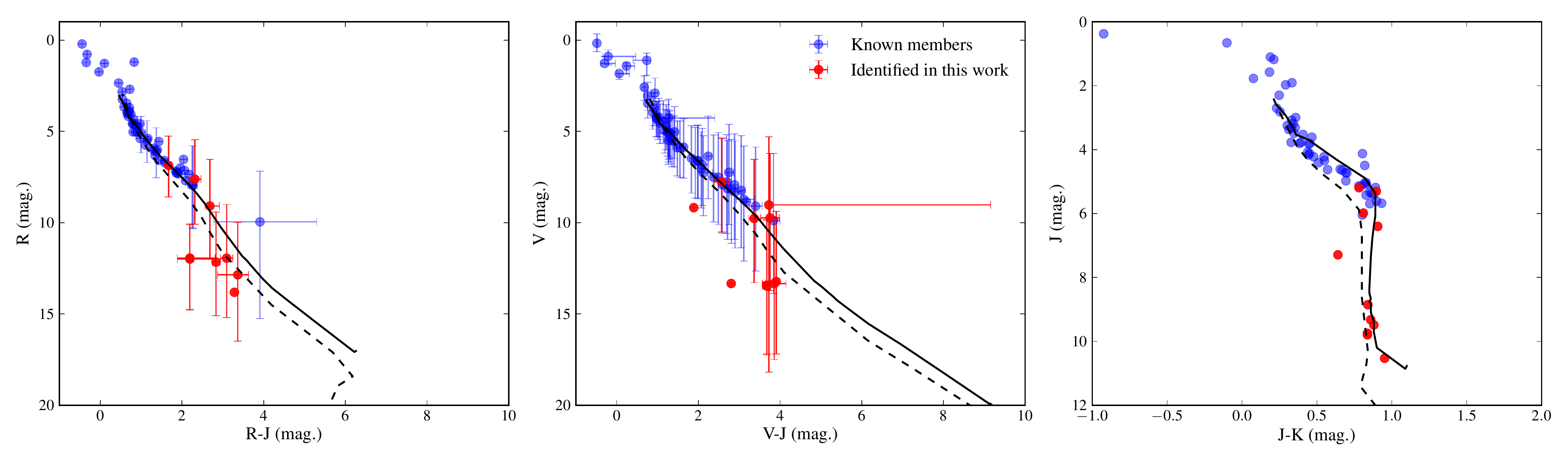}
\vspace{-0.2cm}
\caption{Colour-magnitude diagrams for the Tucana-Horologium  association.}
\label{fig:hr_tha}
\end{center}
\end{figure}

\begin{figure}
\begin{center}
\includegraphics[width=0.95\textwidth]{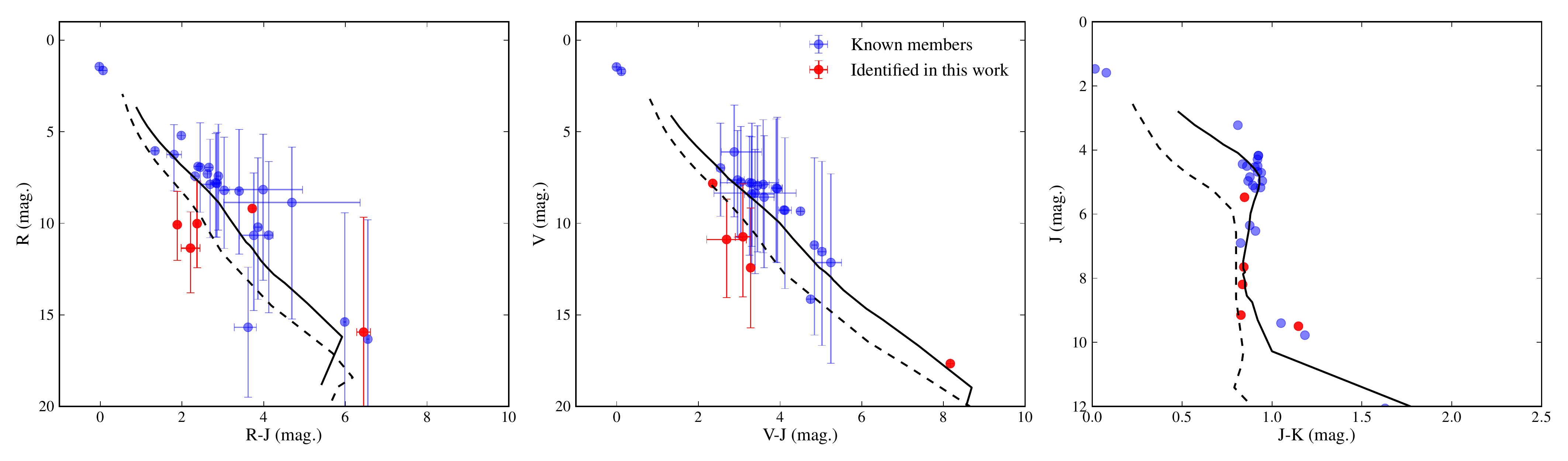}
\vspace{-0.2cm}
\caption{Colour-magnitude diagrams for the TW Hydrae association.}
\label{fig:hr_twa}
\end{center}
\end{figure}

\newpage

\section{Sample used in this work}
\label{sec:orig_sample}

{\onecolumn
\tiny
\LTcapwidth=\textwidth
\begin{longtable}{p{3.8cm} p{1.7cm} p{1.5cm} p{1.0cm}p{0.7cm}p{0.7cm}p{0.7cm} p{1.5cm} p{1.5cm} p{0.7cm} p{0.7cm}}
\caption{Basic properties of the original sample used in this work.}\label{tab:orig_sample}\\
\hline\\
  \multicolumn{1}{l}{ID} &
  \multicolumn{1}{l}{RA} &  
  \multicolumn{1}{l}{DEC} &   
  \multicolumn{1}{l}{Dist.} &      
  \multicolumn{1}{l}{$J$} &  
  \multicolumn{1}{l}{$H$} &    
  \multicolumn{1}{l}{$K$} &
  \multicolumn{1}{l}{$\mu_{\alpha,*}$} &  
  \multicolumn{1}{l}{$\mu_\delta$} &    
  \multicolumn{1}{l}{Mem ref.} \\
  \multicolumn{1}{l}{} &
  \multicolumn{1}{l}{hh:mm:ss.s} &  
  \multicolumn{1}{l}{dd:mm:ss.s} &   
  \multicolumn{1}{l}{(pc)} &      
  \multicolumn{1}{l}{(mag.)} &  
  \multicolumn{1}{l}{(mag.)} &    
  \multicolumn{1}{l}{(mag.)} &
  \multicolumn{1}{l}{(mas/yr)} &  
  \multicolumn{1}{l}{(mas/yr)} &    
  \multicolumn{1}{l}{} \\
  [1ex]
\hline\\
\endfirsthead
\hline\hline\\
  \multicolumn{1}{l}{ID} &
  \multicolumn{1}{l}{RA} &  
  \multicolumn{1}{l}{DEC} &   
  \multicolumn{1}{l}{Dist.} &      
  \multicolumn{1}{l}{$J$} &  
  \multicolumn{1}{l}{$H$} &    
  \multicolumn{1}{l}{$K$} &
  \multicolumn{1}{l}{$\mu_{\alpha, *}$} &  
  \multicolumn{1}{l}{$\mu_\delta$} &    
  \multicolumn{1}{l}{Mem ref.} \\
  \multicolumn{1}{l}{} &
  \multicolumn{1}{l}{hh:mm:ss.s} &  
  \multicolumn{1}{l}{dd:mm:ss.s} &   
  \multicolumn{1}{l}{(pc)} &      
  \multicolumn{1}{l}{(mag.)} &  
  \multicolumn{1}{l}{(mag.)} &    
  \multicolumn{1}{l}{(mag.)} &
  \multicolumn{1}{l}{(mas/yr)} &  
  \multicolumn{1}{l}{(mas/yr)} &    
  \multicolumn{1}{l}{} \\
\hline\\
\endhead
\multicolumn{10}{c}{ABD} \\\hline\\
PW And  &  00:18:20.9  &  +30:57:22.0  &  27.3  &  7.0  &               6.5  &  6.4  &  144.4$\pm$1.0  &  -170.8$\pm$0.9  &  1 \\
V493 And AB  &  00:34:8.4  &  +25:23:50.0  &  49.9  &  8.5  &           7.8  &  7.7  &  81.5$\pm$1.8  &  -97.6$\pm$1.1  &  3 \\
HD 4277  &  00:45:50.9  &  +54:58:40.0  &  52.7\tablefoottext{a}  &  6.6  &              6.4  &  6.4  &  95.6$\pm$1.0  &  -76.9$\pm$1.0  &  1 \\
HIP 4967  &  01:03:40.1  &  +40:51:29.0  &  29.6\tablefoottext{a}  &  8.1  &              7.5  &  7.3  &  126.7$\pm$0.7  &  -167.2$\pm$0.8  &  1 \\
HD 6569  &  01:06:26.2  &  -14:17:47.0  &  47.4\tablefoottext{a}  &  7.9  &              7.4  &  7.3  &  99.8$\pm$1.6  &  -91.8$\pm$2.5  &  1 \\
BD-12 243  &  01:20:32.3  &  -11:28:4.0  &  34.3\tablefoottext{a}  &  7.0  &              6.7  &  6.5  &  112.0$\pm$2.3  &  -137.7$\pm$1.3  &  1 \\
\hline
\end{longtable}
\noindent{\bf \footnotesize Notes. }{\footnotesize $^{(a)}$ Distances from parallax measurements\\}
{\bf \footnotesize References. }{\footnotesize 1: SACY works \citep{Torres2008, Elliott2014}, 2: \cite{Zuckerman2011}, 3: \cite{Malo2014}, 4: \cite{Murphy2015}, 5: \cite{Kraus2014}.
}
}

\section{All properties and references of targets identified in this work}
\label{sec:master_comp_table}

{
\centering
\begin{table}
{
\label{tab:online_appendix_tab}
\begin{tabular}{p{2.3cm}}
\end{tabular}}
\end{table}}

\end{appendix}

\end{document}